\documentclass[a4paper,fleqn,usenatbib]{mnras}
\usepackage[T1]{fontenc}
\usepackage{ae,aecompl}

\usepackage{graphicx}	
\usepackage{amsmath}	
\usepackage{amssymb}	
\usepackage[outdir=./]{epstopdf}
\usepackage{epstopdf}
\usepackage{lscape}
\usepackage{rotating}

\title[Cosmological Discs]{Disc-Halo Interactions in $\Lambda $CDM}

\author[Jacob S. Bauer et al.]{
Jacob S. Bauer,$^{1}$\thanks{E-mail: jacob.bauer@queensu.ca}
Lawrence M. Widrow,$^{1}$
Denis Erkal$^{2\,3}$
\\
$^{1}$Department of Physics, Engineering Physics \& Astronomy, Queen's University, Stirling Hall, Kingston, ON K7L 3N6, Canada\\
$^{2}$Institute of Astronomy, Madingley Road, Cambridge, CB3 0HA, UK\\
$^{3}$Department of Physics, University of Surrey, Guildford, GU2 7XH, UK\\
}

\date{Accepted XXX. Received YYY; in original form ZZZ}

\pubyear{2017}

\begin{document}

\label{firstpage}
\pagerange{\pageref{firstpage}--\pageref{lastpage}}
\maketitle

\begin{abstract}

We present a new method for embedding a stellar disc in a
cosmological dark matter halo and provide a worked example from a
$\Lambda$CDM zoom-in simulation.  The disc is inserted into the halo
at a redshift $z=3$ as a zero-mass rigid body.  Its mass and size
are then increased adiabatically while its position, velocity, and
orientation are determined from rigid-body dynamics.  At $z=1$,
the rigid disc is replaced by an N-body disc whose particles sample
a three-integral distribution function (DF).  The simulation then
proceeds to $z=0$ with live disc and halo particles.  By comparison,
other methods assume one or more of the following: the
centre of the rigid disc during the growth phase is pinned to the
minimum of the halo potential,  the orientation of the rigid disc is
fixed,  or the live N-body disc is constructed from a two rather than
three-integral DF.  In general, the presence of a disc makes the halo
rounder, more centrally concentrated, and smoother, especially in
the innermost regions.  We find that methods in which the disc is
pinned to the minimum of the halo potential tend to overestimate the
amount of adiabatic contraction.  Additionally, the effect of
the disc on the subhalo distribution appears to be rather insensitive
to the disc insertion method.  The live disc in our simulation
develops a bar that is consistent with the bars seen in late-type
spiral galaxies. In addition, particles from the disc are
launched or ``kicked up'' to high galactic latitudes.

\end{abstract}


\section{Introduction}

The structure and evolution of galaxies are determined by the spectrum
of primordial density perturbations, the dynamics of stars and dark
matter, and baryonic physics.  Over the past two decades, there has
been a concerted effort to incorporate the latter into cosmological
simulations \citep[e.g.][]{katz1996feedback, springel2003feedback,
  stinson2006, RoskarDiskMisalignment, pakmorMHD, gomezwarps}.  While
these simulations have enhanced our understanding of galaxy formation,
their computational cost is high.  Adding to the challenge is the
complex and sub-grid nature of star formation, supernova feedback, and
other baryonic processes, which require {\it ad hoc} parametric
models.

In this work, we focus on the dynamics of disc galaxies.  Our goal is
to study the nature of disc-halo interactions where it is advantageous
to be able to control properties of the disc such as its mass, size,
and internal kinematics.  Such control is not possible in {\it ab
  initio} simulations.

Simulations of isolated disc galaxies provide an alternative arena to
study galactic structure and dynamics.  Moreover, many aspects of
disc-halo interactions can be understood by considering the
collisionless dynamics of stars and dark matter while ignoring gas
physics.  For example, simulations of stellar disc-bulge systems
embedded in dark haloes have proved indispensable in studies of bar
and spiral structure formation (See \citet{Sellwood2013} and
references therein).  These simulations typically begin with systems
that are in equilibrium, or nearly so.  For this reason, they usually
assume axisymmetric initial conditions, which are manifestly
artificial.  In short, discs do not come into existence as formed,
highly symmetric objects but rather build up through the combined
effects of gas accretion, star formation, and feedback
\citep{IllustrisFeedback, Eagle}.  Moreover, the haloes in which the
real discs reside are almost certainly triaxial and clumpy
\citep{NFW,mooresubhalos,Klypin1999}.

There now exists a long history of attempts to bridge the gap between
simulations of isolated disc-bulge-halo systems, with their pristine
initial conditions, and cosmological simulations.
\citet{kazantzidis2008}, for example, followed the evolution of a
Milky Way-like disc in its encounter with a series of satellites whose
properties were motivated by cosmological simulations.  They found
that the satellites ``heated'' the disc and prompted the formation of
a bar and spiral structure.  Along similar lines, \citet{purcell2011}
modeled the response of the Milky Way to the gravitational effects of
the Sagittarius dwarf galaxy (Sgr) by simulating disc-satellite
encounters for different choices of the satellite mass.  They
concluded that Sgr may have triggered the development of the spiral
structure seen in the Milky Way today. Continuing in this vein, \citet{laporte_et_al_2016} studied the
influence of the Large Magellanic Cloud and Sgr on the Milky Way disc and
found that they can create similar warps to what has been observed. The effect of a time-dependent triaxial halo
was investigated in \citet{hu_sijacki_2016} where they found it can trigger grand-design spiral
arms. 

Of course, the disc of the Milky Way lives within a population of
satellite galaxies and, quite possibly, pure dark matter subhaloes
\citep{mooresubhalos,Klypin1999}.  With this in mind \citet{Font2001}
simulated the evolution of an isolated disc-bulge-halo model where the
halo was populated by several hundred subhaloes.  They concluded that
that substructure played only a minor role in heating the disc, a
result that would seem at odds with those of \citet{kazantzidis2008}.
Numerical simulations by \citet{gauthier2006} and \citet{dubinski2008}
shed some light on this discrepancy.  In those simulations, 10\% of
the halo mass in an isolated disc-bulge-halo system was replaced by
subhaloes with a mass distribution motivated by the cosmological
studies of \citet{gao2004}.  \citet{gauthier2006} found that a modest
amount of disc heating occurred during the first 5 Gyr, at which point
satellite interactions prompted the formation of a bar, which in turn
heated the disc more significantly.  Not surprisingly, the timing of
bar formation varied from 1 Gyr to 10 Gyr when the experiment was
repeated with different initial conditions for the satellites.

The aforementioned simulations have several drawbacks.  First, most of them do
not allow for halo triaxiality.  Second, the disc is initialized at
its present-day mass whereas real discs form over several Gyr.
Finally, the subhaloes are inserted into the halo in an \textit{ad
  hoc} fashion.  Several attempts have been made to grow a stellar
disc in a cosmological halo in an effort to address these
shortcomings \citep{BerentzenShlosmanStellarDisks,DeBuhrStellarDisks, YurinSpringelStellarDisks}.  The general scheme proceeds in three stages.  During
the first stage, a cosmological simulation is run with pure dark
matter and a suitable halo is selected.  In the second, a rigid disc
potential is grown slowly in the desired halo, thus allowing the halo
particles to respond adiabatically to the disc's time-varying
potential.  In the third stage, the rigid disc is replaced by a live
one and the simulation proceeds with live disc and halo particles.

\citet{DeBuhrStellarDisks} used such a scheme to introduce stellar
discs into dark matter haloes from the Aquarius Project
\citep{springel2008}.  They added a rigid disc at a redshift $z=1.3$
with a mass parameter for the disc that grew linearly with the scale
factor from an initial value of zero to its final value at $z=1$.  The
disc was initially centered on the potential minimum of the halo and
oriented so that its symmetry axis pointed along either the minor or
major axis of the halo.  During the rigid disc phase, the motion of
the disc centre of mass was determined from Newton's 3rd law.
To initialize the live disc, \citet{DeBuhrStellarDisks} approximate
the halo potential as a flattened, axisymmetric logarithmic potential
and then determine the disc distribution function (DF) by solving the
Jeans equations.  

\citet{YurinSpringelStellarDisks} introduced a number of improvements
to this scheme.  Most notably, they use \textsc{galic} to initialize
the live disc \citep{YurinSpringelGalic}.  This code is based on an
iterative scheme for finding stationary solutions to the collisionless
Boltzmann equation.  The general idea for iterative codes is to begin with a set of
particles that has the desired spatial distribution and some initial
guess for the velocity distribution.  The velocities are then adjusted
so as to achieve stationarity, as measured by evolving the system and
computing a certain merit function.  In
\citet{YurinSpringelStellarDisks} the initial disc was assumed to be
axisymmetric with a DF that depended on two integrals of motion, the
energy, $E$, and angular momentum, $L_z$. One striking, if not puzzling,
result from this work is the propensity of the discs to form very
strong bars.  These bars are especially common in models without
bulges even in cases where the disc is submaximal.

In this paper, we introduce an improved scheme for inserting a live
disc in a cosmological halo.  In particular, the centre of mass {\it
  and} orientation of the rigid disc are determined by solving the
standard equations of rigid body dynamics.  Thus, our rigid disc can
undergo precession and nutation.  The angular and linear velocities of
the rigid disc at the end of the growth phase are incorporated into
the live disc initial conditions.  As in
\citet{YurinSpringelStellarDisks} we use an axisymmetric approximation
for the halo potential when constructing the disc DF.  However, our DF
is constructed from an analytic function of $E$, $L_z$, and the
vertical energy $E_z$, which is an approximate integral of motion used
in \textsc{galactics} \citep{DubinskiKuijkenRigidDisks,
  WPDGalactICSReference}.  By design, the disc DF yields a model whose
density has the exponential-${\rm sech}^2$ form.  And with a
three-integral DF, we have sufficient flexibility to model realistic
Milky Way-like discs.  As discussed below, the initial disc DF may be
crucial in understanding the formation of the bar.

As a demonstration of our method we grow a Milky Way-like disc in an approximately
$10^{12}\,h^{-1}\,M_\odot$ halo from a cosmological zoom-in
simulation.  We discuss both disc dynamics and the effect our disc has
on the population of subhaloes.  Discs have been invoked as a means of
depleting halo substructure and thus alleviating the Missing Satellite
Problem, which refers to the underabundance of observed Milky Way
satellites relative to the number of Cold Dark Matter subhaloes seen
in simulations \citep{mooresubhalos,Klypin1999}.  An earlier study by
\citet{DOhngiaSubstructureDepletion} found that when a disc potential
is grown in a Milky Way-size cosmological halo, the abundance of
substructure in the mass range $10^7\,M_\odot$ to $10^9\,M_\odot$ was
reduced by a factor of $2-3$.  Similar results were found by
\citet{Sawala2017} and \citet{GKSubhaloDepletion17}.

The organization of the paper is as follows.  In Section 2, we outline
our method for inserting a live disc into a cosmological simulation.
We also present results from a test-bed simulation where a disc is
inserted into an isolated flattened halo.  We next apply our method to
a cosmological zoom-in simulation.  In Section 3, we focus on disc dynamics
and find that the disc develops a bar, spiral structure and a warp.
In addition, disc-halo interactions appear to ``kick'' stars out of
the disc and into regions normally associated with the stellar halo.
In Section 4, we present our results for the spherically-averaged
density profile and shape of the dark matter halo as well as the
distribution of subhaloes.  Particular attention is paid to the
sensitivity of these results to the disc insertion scheme.  We
conclude in Section 5 with a summary and discuss possible applications
of this work.

\begin{table*}
\begin{tabular}{l l l l l l l}
\hline
 & DMO & MN & FO & RD & LD\\
\hline
$M_d \, (M_\odot)$ & -- & $7.2 \times 10^{10}$ & $7.2 \times 10^{10}$ & $7.2 \times 10^{10}$ & $7.2 \times 10^{10}$\\
$R_{d,0}$ (kpc) & -- & 3.7 & 3.7 & 3.7 & 3.7\\
$N_d$ & -- & -- & $10^6$& $10^6$ & $10^6$\\
$z_g$ & -- & 3.0 & 3.0 & 3.0 & --\\
$z_l$ & -- & 1.0 & 1.0 & 1.0 & 1.0\\
$N_r$ & 4096 & 4096 & 4096 & 4096 & 4096\\
$L_{box} (\text{ Mpc} \,h^{-1} )$ & 50  & 50 & 50 & 50 & 50\\
\hline
\end{tabular}\caption{A summary of the simulation parameters, as discussed in the text. $M_d$ is the final disk mass, $R_{d,0}$ is the final disk scale radius, $N_d$ is the number of particles used to simulate the disk, $z_g$ and $z_l$ are the redshifts when the disk beings to grow and when it becomes live (respectively), $N_r$ is the effective resolution in the zoom-in region, and $L_{box}$ is the comoving size of the box. } \label{tab:simparams}
\end{table*}

\section{Inserting a Stellar Disc into a Cosmological Halo}

In this section, we detail our method for inserting a live stellar
disc into a cosmological simulation.  We begin with an overview of our
approach and the five main simulations presented in this paper.  We
then describe some of the more technical aspects of the method.

\subsection{Overview of Simulation Set} \label{sec:sim_overview} 

Our simulations are performed with the N-body component of
\textsc{gadget-3}, which is an updated version of \textsc{gadget-2}
\citep{springel_2005}.  For the cosmological simulations, we implement
the zoom-in technique of \citet{KatzQuasarZoom} and
\citet{NavarroWhiteZoom}, broadly following the recommendations of
\cite{onorbe_etal_2014}, which allows us to achieve very high spatial
and mass resolution for a single halo while still accounting for the
effects of large-scale tidal fields. For the cosmological parameters, we
use the results from Planck 2013 \citep{planck_2014} with $h=0.679$, $\Omega_b = 0.0481$, 
$\Omega_0 = 0.306$, $\Omega_\Lambda = 0.694$, $\sigma_8 = 0.827$, and $n_s = 0.962$. 

We begin by simulating a $50\,h^{-1} {\rm Mpc}$ box with $N_r=512^3$ particles, where $N_r$ is the effective resolution, each with a mass 
of $\sim 7.9\times 10^{7}\,h^{-1} M_\odot$.  We identify a Milky Way-like
halo in the present-day snapshot, that is, a $\sim 10^{12}\,M_\odot$
halo which has experienced no major mergers since $z=1$ and which has
no haloes with $2 h^{-1}$ Mpc more than half the mass of the MW-analogue. We then run an
intermediate zoom-in simulation targeting all particles within 10
virial radii of the low resolution halo.  The initial conditions for
this simulation are generated with \textsc{music} \citep{music}, which
creates five nested regions from a coarse resolution of $N_r=128^3$ in the
outskirts to an effective $N_r=2048^3$ resolution in the targeted region.
After this simulation reaches $z=0$, we select all particles within
7.5 virial radii and regenerate initial conditions with one more level
of refinement, giving six nested zoom regions, where now, the highest
effective resolution is $N_r=4096^3$. Our final halo is composed of
approximately $10^7$ particles, each with a mass of
$1.54\times 10^5\,h^{-1} M_\odot$. The softening lengths were selected using
the criteria in \citet{power_et_al_2003} with a softening of $719$ comoving pc for the highest resolution 
particles in the final zoom-in simulation. We found that this repeated zoom-in technique
results in remarkably little contamination from coarse resolution
particles within the targeted halo, giving a clean region of size $1.9 h^{-1}$ Mpc at $z=0$.

The dark matter only (DMO) simulation not only serves as the basis for
four simulations with discs but also provides a control ``experiment''
for our study of the effect discs have on halo properties.  In each of
our disc simulations, the potential of a rigid disc is introduced at
the ``growing disc'' redshift $z_g$.  The mass parameter of the disc
is then increased linearly with the scale factor
$a = 1/\left (1+z\right )$ from zero to its final value $M_d$ at the
``live disc'' redshift $z_l$.  As described in
Sec.~\ref{sec:rigid_disks}, the radial and vertical scale lengths of
the disc are also increased between $z_g$ and $z_l$ to account for the
fact that discs grow in scale as well as mass while they are being
assembled.

In the first of our disc-halo simulations, dubbed MN, we introduce a
rigid Miyamoto-Nagai disc \citep{MiyamotoNagai}, whose gravitational
potential is given by

\begin{equation}
  \Phi\left (R,\,z\right ) = -\frac{G M_d} {\left (R^2 + \left(R_d +
    \left(z^2 + z_d^2 \right)^{1/2}\right)^{2}\right )^{1/2}}~.
\end{equation}

\noindent For this simulation, which was meant to mimic the scheme in
\citet{DOhngiaSubstructureDepletion}, we assume that the centre of the
disc tracks the minimum of the halo potential while the orientation of
the disc is fixed to be aligned with the $z$-axis of the simulation
box. Note that this is effectively a random direction for the halo.

In the remaining three disc simulations, we grow an
exponential-sech$^2$ rigid disc potential in our halo with mass and
scale-length parameters that grow in time.  For our fixed-orientation
(FO) simulation, the position and velocity of the disc's centre of
mass are determined from Newtonian dynamics while the spin axis of the
disc is initially aligned with the minor axis of the halo at $z = z_g$
and kept fixed in the simulation box frame thereafter.  In this
respect, the simulation is similar to the ones presented in
\citet{DeBuhrStellarDisks} and \citet{YurinSpringelStellarDisks}.  For
the rigid-disc (RD) simulation the disc's orientation, which is now a
function of time, is determined from Euler's rigid body equations.

In the MN, FO, and RD runs, we continue the simulation to the present
epoch ($z=0$) with the assumed rigid disc potential where the mass and
scale length parameters are held fixed and the position and
orientation are calculated as they were during the growth phase.  For
our final live disc (LD) simulation, we swap a live disc for the RD
disc at $z_l$.  Thus, the RD and LD simulations are identical prior to
$z_l$.  All of our discs have a final mass of
$M_d = 7.2 \times 10^{10} M_\odot$, a final scale radius of
$3.7\,{\rm kpc}$, and a final scale height of $0.44\, {\rm kpc}$. Our simulation
parameters can be found summarized in Table \ref{tab:simparams}.

\vspace{0.1in}
\subsection{Summary of Live Disc Insertion
  Scheme} \label{sec:method_outline}

The first step in our disc insertion scheme is to calculate an
axisymmetric approximation to the gravitational potential of the DMO
halo at $z=0$.  We do so using an expansion in Legendre polynomials as
described below.  We then generate a particle representation of a
stellar disc that is in equilibrium with this potential using the
\textsc{galactics} code
\citep{KGGalactICSReference,WPDGalactICSReference}.  Though
\textsc{galactics} allows one to generate the phase space coordinates
of the disc stars, at this stage, we only need the positions of the
stars, which we use to represent the ``rigid disc''.  At the $z_g$
snapshot, we incorporate the disc particles into the mass distribution
of the system with the disc centered on the potential minimum of the
halo.  We then rerun the simulation from $z_g$ to $z_l$ with the
following provisos.  First, the mass of the disc is increased linearly
with $a$ from zero to its final value.  Second, the size of the disc
increases with time, which we account for by having the positions of
the disc particles, as measured in the disc frame, expand with time to
their final values at $z_l$.  Finally, the center of mass and
orientation of the disc are determined by integrating the equations of
rigid-body dynamics.  At redshift $z_l$, the DF of the disc is
recalculated assuming the same structural parameters as before but
with an axisymmetric approximation to the new halo potential.  An
N-body disc is generated from this DF and the simulation proceeds with
live disc particles.  In this paper, we choose $z_g = 3$ and $z_l = 1$
so that the growth period lasts from $2.2\,{\rm Gyr}$ to
$5.9\,{\rm Gyr}$ after the Big Bang.  This period in time roughly
corresponds to the epoch of peak star formation in Milky Way-like
galaxies \citep[e.g.][]{van_dokkum_etal_2013}.

Our simulations during this epoch can be used to study the effect of a
disc potential on the evolution of substructure.  On the other hand,
our simulations of the live disc epoch ($z_l > z > 0$) can also be used to
study disc dynamics in a cosmologically-motivated dark halo.

\subsection{Halo Potential} \label{sec:ext_disk_pot}

Our method requires an axisymmetric approximation to the halo
potential centred on the disc.  This approximation is found using an
expansion in spherical harmonics (see \citet{BT}) where only the $m=0$
terms are included.  The potential is then expressed as an expansion
in Legendre polynomials.  We divide the region that surrounds the disc
into spherical shells and calculate the quantities

\begin{equation} \label{eq:ml}
m_{l,i} = \sum_{n\in S_i} m_n P_l(\cos{\theta_n}) ,
\end{equation}

\noindent where the sum is over the halo particles of mass $m_n$ in
the $i$'th shell ($S_i$), $P_l$ are the Legendre polynomials, and
$\left (r,\,\theta,\,\phi\right )$ are spherical polar coordinates
centred on the disc.  The axisymmetric approximation to the potential
is then

\begin{equation}
\Phi_h\left (r,\,\theta\right ) = \sum_{l=0}^\infty A_l\left (r\right
) P_l\left (\cos{\theta}\right )
\end{equation}

\noindent where

\begin{equation}
A_l(r) = -G\left (
\frac{1}{r^l}\int_0^r dr' r'^{l+2} m_l(r')
+r^{l+1}\int_0^r dr' r'^{1-l} m_l(r')\right )
\end{equation}
and $m_l$ is given by Eq. \eqref{eq:ml} for sufficiently small radial
bins.

\subsection{Disc DFs} \label{sec:live_ics}

Armed with an axisymmetric approximation to the halo potential, we
construct a self-consistent DF for the disc following the method
outlined in \citet{DubinskiKuijkenRigidDisks}.  This DF is an analytic
function of $E$, $L_z$, and $E_z$.  By construction, the DF yields a
density law for the disc which is, to a good approximation, given by

\begin{equation}
\rho_d\left (R,\,z\right )\simeq \frac{M_d}{4\pi R_d^2 z_d} e^{-R/R_d}
    {\rm sech}^2\left (z/z_d\right ) T\left (R_t,\Delta_t\right )
\end{equation}

\noindent where $M_d$, $R_d$, and $z_d$ are the mass, radial scale
length and vertical scale height of the disc and $R=\sqrt{r^2-z^2}$.
The truncation function $T$ insures that the density falls rapidly to
zero at a radius $R_t + \Delta_t$.  The square of the radial velocity
dispersion is chosen to be proportional to the surface density, that
is, $\sigma_R = \sigma_{R0}\exp{\left (-R/2R_d\right )}$.  The central
velocity dispersion $\sigma_{R0}$ controls, among other things, the
Toomre $Q$ parameter and thus the susceptibility of the disc to
instabilities in the disc plane.  The azimuthal velocity dispersion is
calculated from the radial velocity dispersion through the epicycle
approximation \citep[for details see][]{BT} while the vertical velocity
dispersion is adjusted to yield a constant scale height $z_d$.  We
stress that although the density law is written as a function of $R$
and $z$, which are not integrals of motion, the underlying DF is a
function of $E$, $L_z$, and $E_z$, which are integrals of motion to
the extent that the epicycle approximation is valid and that the
potential can be approximated by an axisymmetric function.

\subsection{Rigid Disc Dynamics} \label{sec:rigid_disks}

During the disc growth phase, the disc mass is given by

\begin{equation}\label{eq:mass}
M(a) = M_d \left( \frac{a - a_g}{a_l - a_g} \right)~,
\end{equation}

\noindent where $a_g$ is the scale factor evaluated at $z_g$. The
positions of the disc particles expand self-similarly in disc or body
coordinates.  That is, the comoving position of a disc particle in
body coordinates is given by ${\bf s}_i(a) = b(a){\bf s}_i
(a_l )$ where ${\bf s}_i(a)$ is the comoving position
of the $i$'th disc particle in the body frame
\begin{equation}\label{eq:scale}
b(a) = b_g + \left (1 - b_g\right ) \left( \frac{a - a_g}{a_l - a_g} \right)~,
\end{equation}
where $b_g = b(a_g)$, an we choose $b_g = 0.1$. The angular velocity of the disc is described by the vector
$\boldsymbol{\omega} = \left (\omega_x,\, \omega_y,\,\omega_z\right )$
where $\omega_z$ corresponds to the spin of the disc about its
symmetry axis.  We assume that
\begin{equation}\label{eq:omega3}
\omega_z(a) = \omega_z(a_l)\left (\frac{M(a)}{M_d}\right )^{1/\alpha}\,b(a)~,
\end{equation}
\noindent which insures that the disc tracks the Tully-Fisher
relation, $M_d\propto V_d^\alpha\propto \left (\omega_3R_d\right
)^\alpha$ \citep{TullyFisherModern}.  In this work we set
$\alpha=3.5$.

The orientation of the disc is described by its Euler angles.  We
follow the convention of \citet{ThorntonAndMarion} and use
$\phi,\,\theta,\,$ and $\psi$ where the matrix

\begin{equation}
{\cal R} = {\cal R}_z(\phi) {\cal R}_y(\theta) {\cal R}_z(\psi)
\end{equation}

\noindent describes the transformation from the disc body frame to the
simulation frame.  Here ${\cal R}_i(\alpha)$ is the matrix for a
rotation by angle $\alpha$ about the $i$'th axis.  Physically, changes
in $\phi$ and $\theta$ correspond to precession and nutation,
respectively while $\psi$ is a degenerate rotation about the disc's
symmetry axis. The equations of motion for the Euler angles and
angular velocity of the disc, which must account for the
time-dependence of the disc's moment of inertia as well as the fact
that \textsc{gadget-3} uses comoving coordinates, are derived in
Appendix~\ref{sec:derivation}.  These equations allow us to solve for
the orientation of the disc under the influence of torque due to dark
matter.

At the end of the growth phase, we initialize the disc with a DF that
is recalculated using \textsc{galactics}.  As we will see, during the
growth phase the motion of the disc involves a mix of precession and
nutation.  In general, a live disc is not able to support the sort of
rapid nutation seen in the rigid disc, essentially because different
parts of the disc respond to torques from the halo and the disc itself
differently.  We therefore initialize the live disc with an
orientation and precessional motion given by a fixed-window moving average of the
rigid disc coordinates.

\subsection{Test-bed Simulation of an Isolated Galaxy}

We test our method by growing a stellar disc in an isolated, flattened
halo in a non-cosmological simulation.  To initialize the flattened halo we first generate a particle
realization of a truncated, spherically symmetric NFW halo \citep{NFW}
whose density profile is given by

\begin{equation}
\rho(r) = \frac{v_h^2 a_h}{4\pi G} \frac{1}{r\left (r + a_h\right )^2}
\end{equation}
with $a_h = 8\,{\rm kpc}$ and $v_h = 400\,{\rm km\,s^{-1}}$.  The halo
is truncated at a radius much larger than the radius of the disc.  The
$z$ and $v_z$ coordinates are then reduced by a factor of two and the
system is evolved until it reaches approximate equilibrium.  The
result is an oblate halo with an axis ratio of $\sim 0.8$ and a
symmetry axis that coincides with the $z$-axis of the simulation box.
We next grow a rigid disc over a period of $1\,{\rm Gyr}$ to a final
mass of $4.9 \times 10^{10}\,M_\odot$ and final radial and vertical
scale lengths of $R_d =2.5\,{\rm kpc}$ and $z_d = 200\,{\rm pc}$,
respectively.  The disc is grown at an incline of $30^\circ$ relative
to the symmetry plane of the halo.  Doing so allows us to check the
rigid body integration scheme for a case when the symmetry axes of the
disc and halo are initially misaligned.  At $t=1\,{\rm Gyr}$ we
replace the rigid disc with a live one and evolve the system for an
additional 1 ${\rm Gyr}$.  In a separate simulation, we also follow
the evolution of the rigid disc over the same time period, allowing us
to compare the evolution of the live and rigid disc.

\begin{figure}
\includegraphics[width=0.5\textwidth]{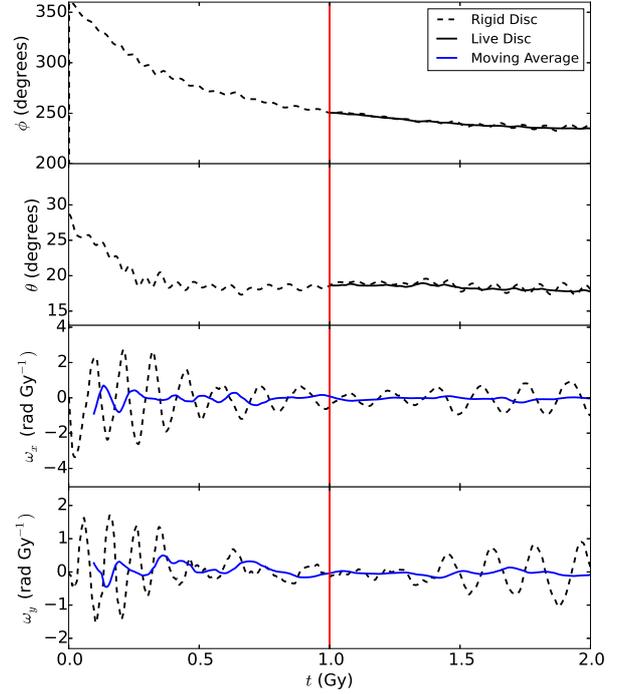}
\caption{Kinematic variables for the rigid and live discs in an
  isolated, flattened halo as a function of time.  The upper two
  panels show the Euler angles $\theta$ and $\phi$ for the rigid disc
  (dashed curves) and live disc (solid curves) where the live disc is
  introduced at $t=1\,{\rm Gyr}$ (red vertical line).  The bottom two
  panels show the $x$ and $y$ components of the angular velocity, as
  measured in the body coordinate system.  In these two panels the
  solid curves show the $\delta t \sim 150\, \text{My}$ moving
  average, which is used to initialize the live disc.}
\label{fig:flattened_halo_regression}
\end{figure}
  
Fig. \ref{fig:flattened_halo_regression} shows the Euler angles and
angular velocity components for the rigid and live discs as a function
of time.  The time-dependence of $\omega_x$ and $\omega_y$ is
characterized by an interference pattern between short $\sim 125\,{\rm
  Myr}$ period nutations and a decaying long-period
precessional motion. Note that $\theta$ and $\phi$ for the live disc
track the corresponding values for the rigid disc for $t>1\,{\rm
  Gyr}$.  By initializing the live disc with the angular velocity of
the rigid disc, we capture the (small) precessional motion of $\sim
10^\circ {\rm Gyr}^{-1}$ between $t = 1\,{\rm Gyr}$ and $2\,{\rm
  Gyr}$.  The disc settles into a preferred plane within the first
$300\,{\rm Myr}$ that is intermediate between its initial symmetry
plane and the initial symmetry plane of the halo. More precisely, the
vector of the new minor axis is $\textbf{c} = [-0.159,0.146,0.976]$
measured at $20\text{ kpc}$, $12.5^{\circ}$ from the original
flattening axis.

\begin{figure*}
\centering 
\includegraphics[width=0.9\textwidth]{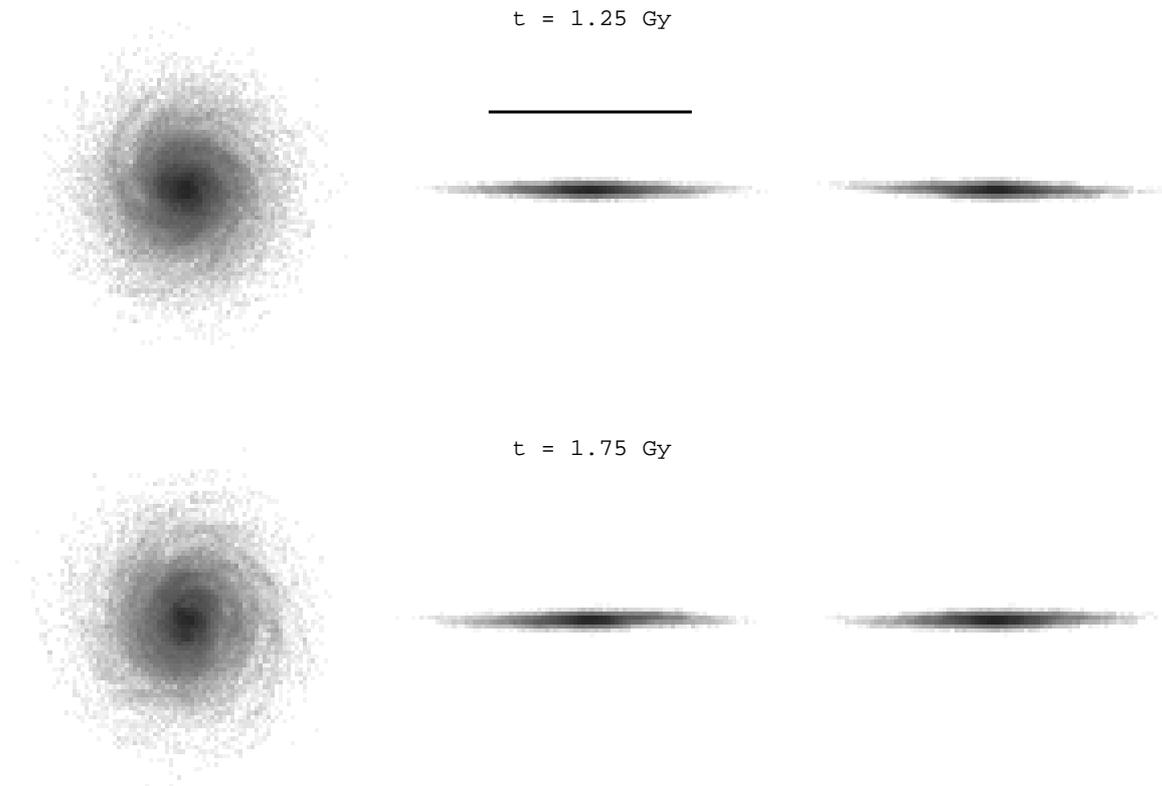}
\caption{Face-on projections of the particle distribution for two
  snapshots of a live disc in a flattened
  halo. The solid line for scale is 25 kpc with a centre
  coincident with the disc's. }\label{fig:flattened_disk_warps}
\end{figure*}

\begin{figure}
\includegraphics[width=0.5\textwidth]{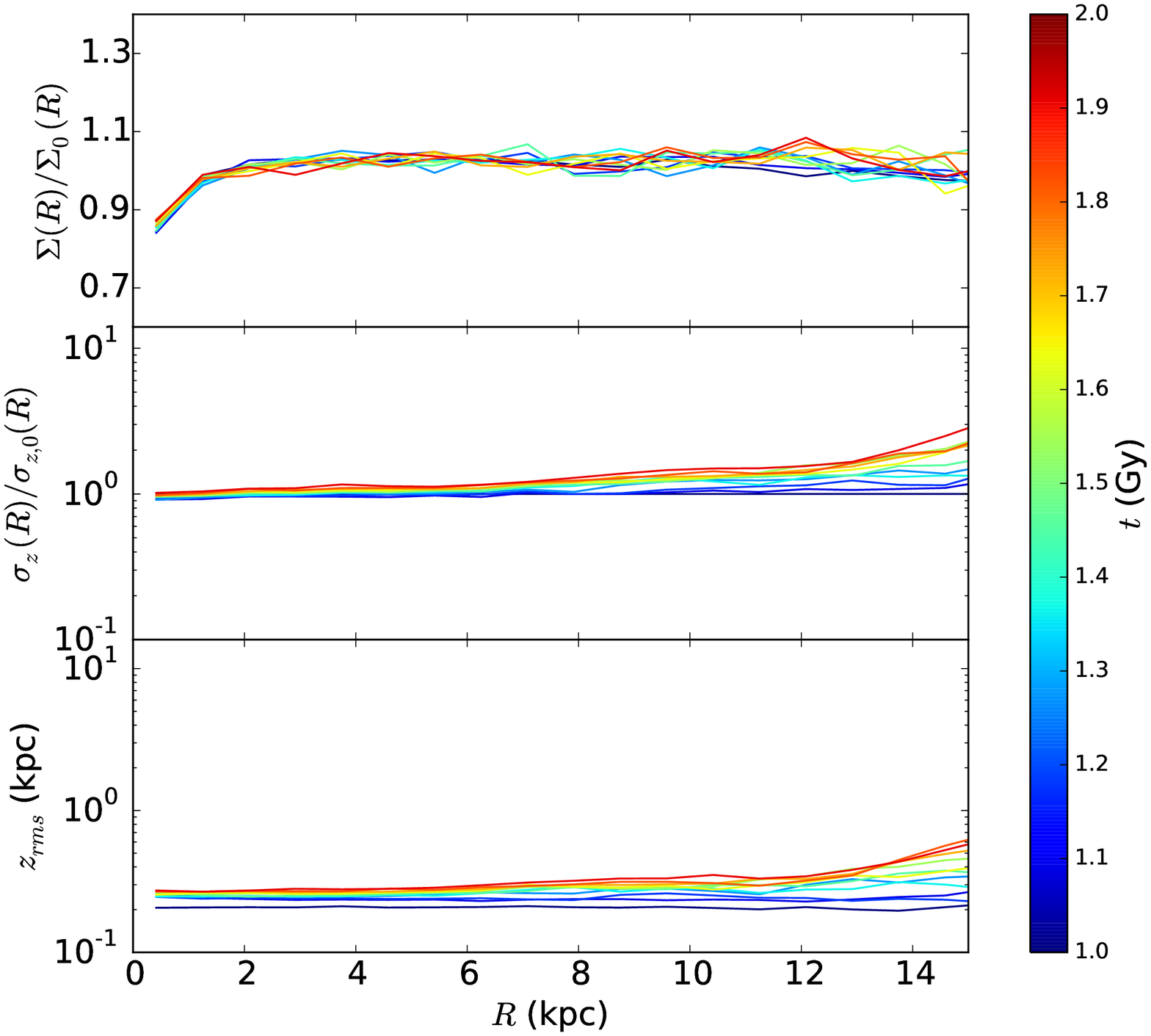}
\caption{Surface density, vertical velocity dispersion, and scale
  height profiles as a function of Galactocentric radius $R$ for 10
  snapshots equally spaced in time.  The top panel shows the surface
  density $\Sigma(R)$ divided by
  $\Sigma_0(R) = \left (M_d/2\pi R_d^2\right )\exp{\left[-\left
        (R/R_d\right ) \right]}$
  in order to highlight departures from a pure exponential disc.
  Likewise, in the middle panel, we show the ratio
  $\sigma_z(R)/\sigma_{z,0}(R)$ where $\sigma_{z,0} = \exp{(-R/2R_d)}$.
  The bottom panel shows the RMS $z$ as a function of $R$. }
\label{fig:flattened_surface_densities}
\end{figure}

Fig. \ref{fig:flattened_disk_warps} shows surface density maps for the
disc at two snapshots.  The disc develops a weak warp due to its
interaction with the halo.  The development of the warp is also
evident in the surface density, vertical velocity dispersion, and
scale height profiles shown in
Fig. \ref{fig:flattened_surface_densities}.  We see that the surface
density within $\sim 15\,{\rm kpc}$ or $6R_d$ is essentially unchanged
while at larger radii, there are $10-20\%$ time-dependent
fluctuations.  The scale height $\langle z^2\rangle^{1/2}$ increases
with time and radius.  At early times, the increase is most prominent
beyond $\sim 15\,{\rm kpc}$ while at late times, the scale height
increases more smoothly from the center to the edge of the disc.

\section{Cosmological Simulations}

We now use our method to insert a live disc with prescribed structural 
properties into a cosmological halo.  In this section, we focus on 
disc dynamics while in the next, we consider the effect the disc has 
on the dark halo. 

\begin{figure}
\includegraphics[width=0.5\textwidth]{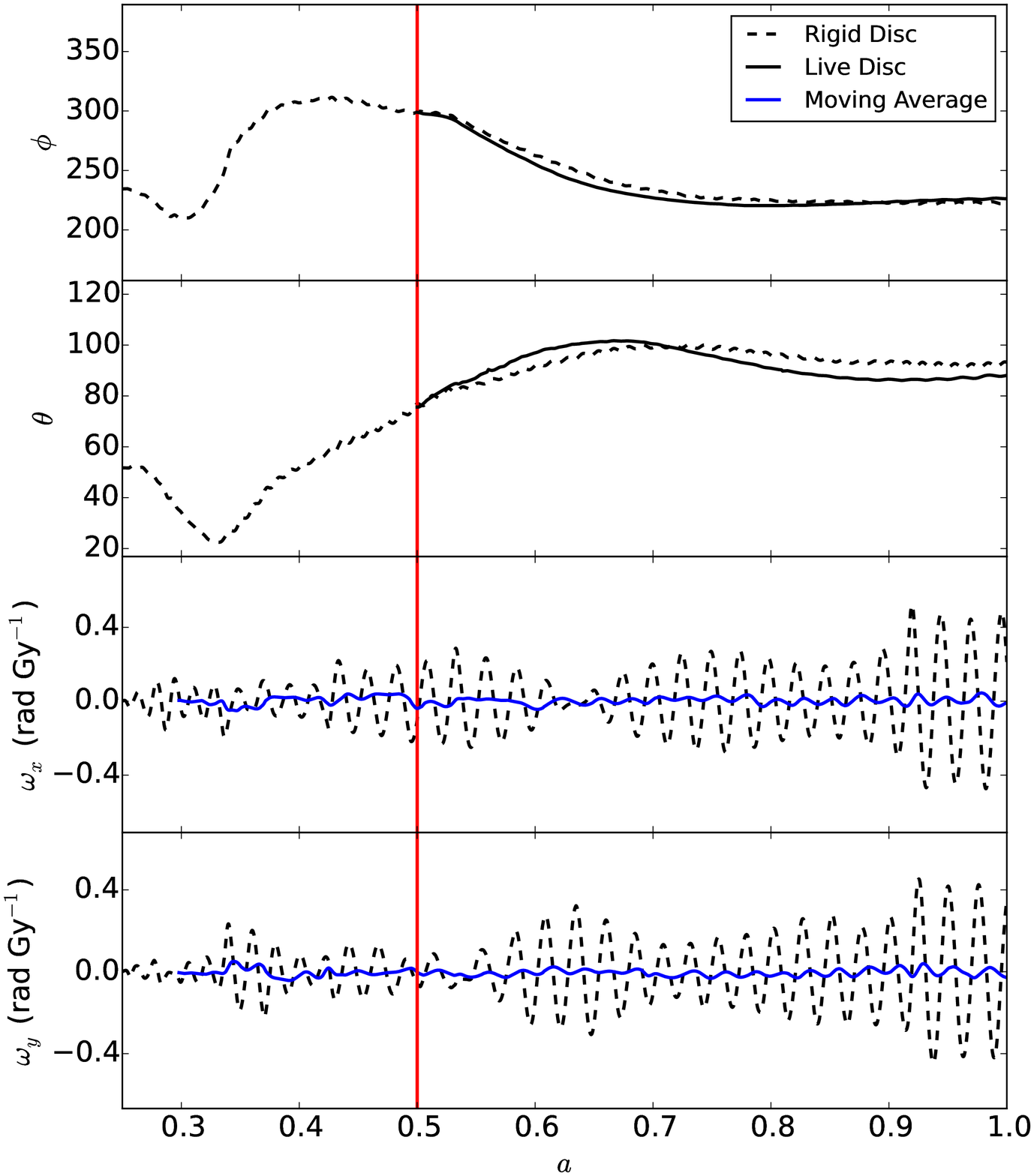}
\caption{Kinematic variables for the rigid and live discs in our
  cosmological halo as a function of scale factor $a$.  Line types are
  the same as in Fig. \ref{fig:flattened_halo_regression}.  The live
  disc is introduced at a redshift $z=1$ when the scale factor is
  $a=0.5$ (red vertical line). The blue line shows the $\delta a \sim
  0.04$ moving average calculated by averaging the last 300 points in
  the disc integration routine.} \label{fig:cosmo_inclination}
\end{figure}

In Fig.\,\ref{fig:cosmo_inclination} we show the kinematic variables
for the rigid and live discs in the RD and LD simulations.  The two
simulations are identical prior to $z=1$ ($a=0.5$) when the live disc
is swapped in for the rigid one.  The short period ($300\,{\rm Myr}$)
oscillations in $\omega_1$ and $\omega_2$ are nutations.  To
initialize the live disc, we use the fixed-window moving average of $\omega_x$ and
$\omega_y$.  By and large, the Euler angles of the rigid and live
disc's track one another for $z<1$, indicating that the rigid disc
is a reasonable model for a live one, at least in terms of the disc's
orientation.

\begin{figure*}
\centering 
\includegraphics[width=1.\textwidth]{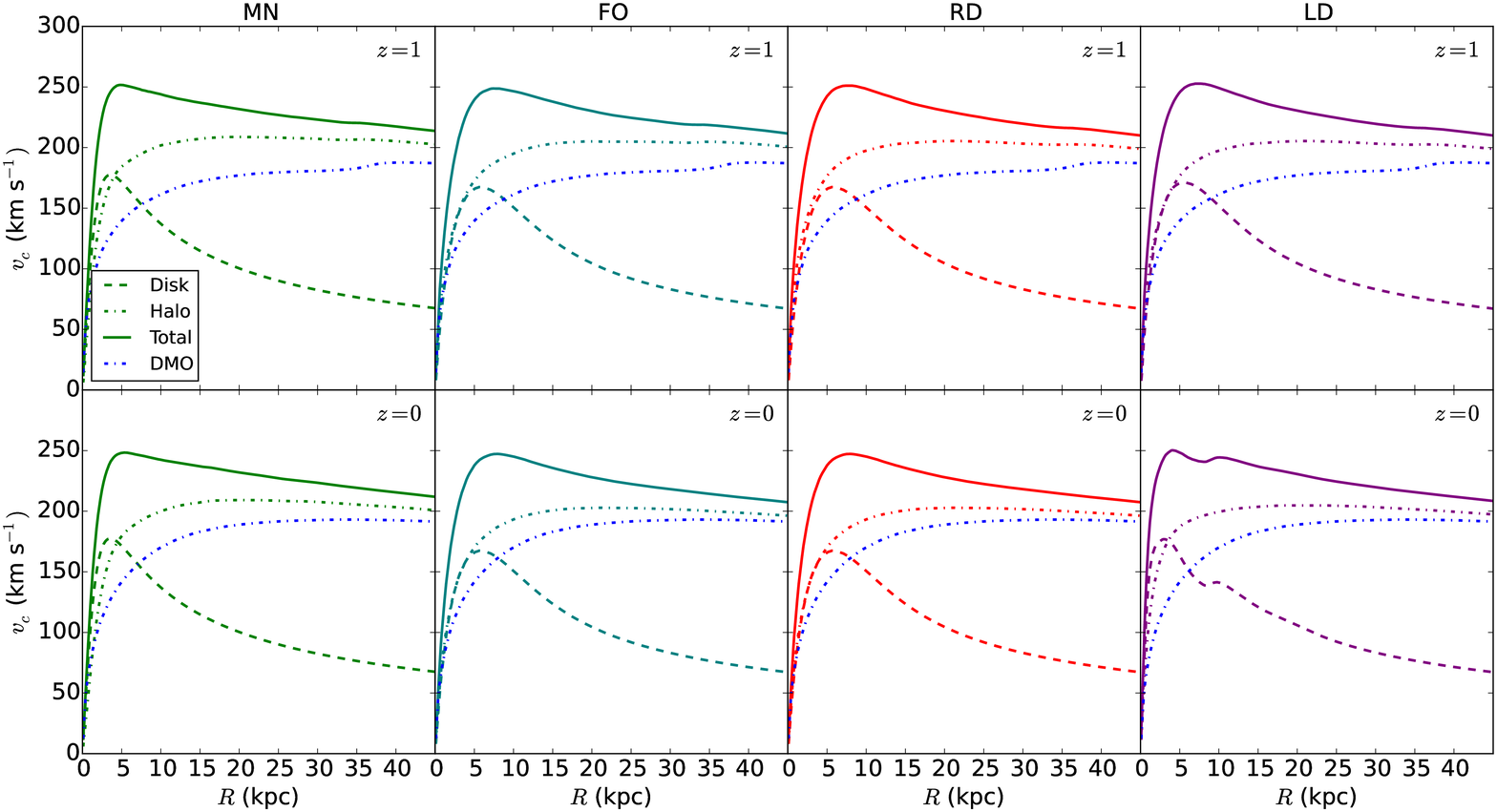}
\caption{Circular speed curve decompositions at $z=1$ (top row) and
  $z=0$ (bottom row) for (from left to right) our LD, RD, and MN
  simulations.  Halo contributions are represented as dot-dashed
  lines, disc contributions are represented by dashed lines, and the
  total rotation curve is given by a solid line.  For reference, we have
  included the circular speed curve for the DMO halo (dot-dashed curve).}
\label{fig:rotation_curves}
\end{figure*} 

In Fig.\,\ref{fig:rotation_curves} we show the circular speed curves
at $z=1$ and $z=0$ for our four simulations.  We see that the disc in
our model is submaximal.  To be precise, we have $V_d/V_c \simeq 0.68$
at $R=2.2R_d$ where $V_d$ is the circular speed due to the disc and
$V_c$ is the total circular speed.  In short, the contributions from
the disc and halo to the centrifugal force are approximately equal at
a radius where the disc contribution reaches its peak value.  By
comparison, a maximal disc is generally defined to have $V_d/V_c >
0.85$ \citep{sackett1997}.  If we use $V_d/V_c$ at $2.2R_d$ as the
defining characteristic of the model, then our simulations match up
with the F-5 simulation of
\citet{YurinSpringelStellarDisks}, although our discs are slightly
warmer, with a Toomre Q-parameter of $1.4$ as compared with $Q\simeq
0.9$ for their discs and our discs are thinner ($200\,{\rm pc}$ vs.
$600\,{\rm pc}$). We note that in a two-integral disc DF, $Q$ and the
disc thickness are linked whereas in a three-integral DF, they can be
set independently.  The method of \citet{YurinSpringelGalic} can be extended to consider
a three-integral DF, but these models were not considered in  \citet{YurinSpringelStellarDisks}. Moreover, their two-integral model, 
which imposes $\sigma_R=\sigma_z$, violates the epicycle approximation,
leading to transient system behaviour at early times when disc bars
first form.


The circular speed curves in Fig. \ref{fig:rotation_curves} show little change exterior to $\sim 2R_d$
after $z_l$, thus providing another indication that the live disc
was close equilibrium when it was swapped in for the rigid one. The
formation of a bar is evident in the circular speed, and we can infer
the bar contributes substantially inside $2.2 R_d \simeq 8$\,kpc.  The halo contribution at $R=2.2R_d$ is about 20\% larger
in the four disc runs than in the DMO one due to adiabatic
contraction.  Interestingly, the halo in the MN run shows
somewhat more contraction than in the RD and LD runs.  We note that in
the MN run, the disc potential tracks the potential minimum of the
halo whereas in the RD/LD case, the disc's position is determined from
Newtonian dynamics.  In general, the centre of the disc tracks the
halo potential minimum so long as the potential changes slowly with
time.  However, during a major merger (and indeed, just such an event
occurs at $z=2$) there are rapid changes in the halo potential and 
the position of the disc, as determined by Newtonian dynamics, can
differ significant from the minimum of the halo potential.  Evidently,
the {\it ad hoc} prescription of growing a disc at the halo's
potential minimum may, in some cases, over-estimate the effect of
adiabatic contraction.

\begin{figure*}
\includegraphics[width=0.9\textwidth]{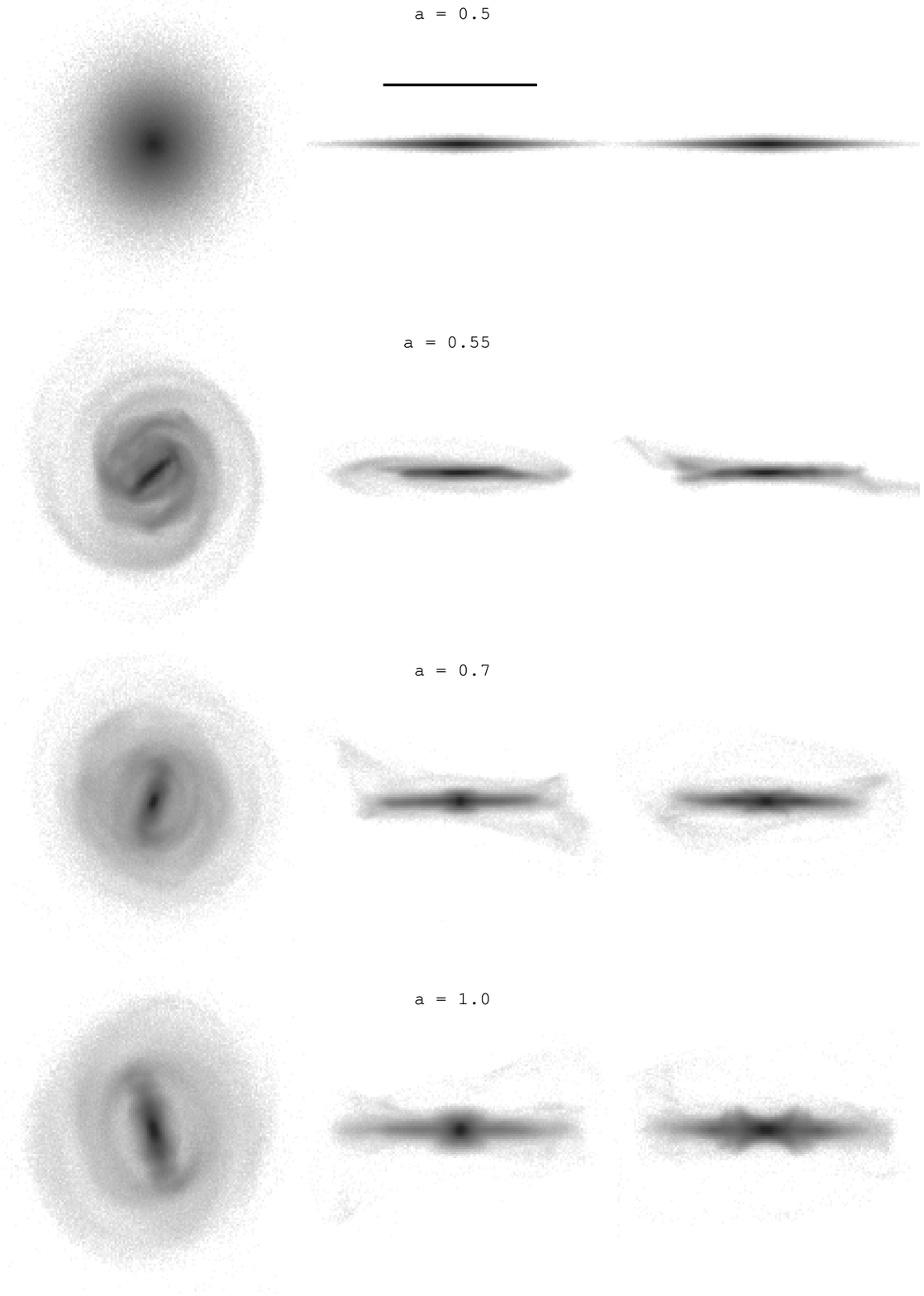}
\caption{Projected density along three orthogonal directions for the
  live disc at four epochs between $z=1$ and $z=0$.  The projections are presented
  in physical units. The solid line for scale is
  37 kpc with a centre coincident with the disc's.}
\label{fig:cosmo_density_panels}
\end{figure*}

\subsection{Bar Formation}

In Fig.\,\ref{fig:cosmo_density_panels} we show orthogonal projections
of the disc density in our LD simulation at four epochs between
$z=1$ (lookback time of 7.9 Gyr) and the present epoch.
During the first billion years of live disc evolution, the disc
develops a bar and spiral structure.  In addition, there is a
warp in the outer disc extending several kiloparsecs above the midplane
of the inner disc.  By the present epoch, the bar has grown in length and
intensified and the edge-on view shows the classic X-pattern.

We consider the usual parameter bar strength $A_2 = 
|c_2|$ where

\begin{equation}
c_m = \frac{1}{M_S}\sum_{j\in S} m_j e^{im\phi_j}~.
\end{equation}

\noindent Here, $S$ is some circularly-symmetric region of the disc
(e.g., a circular annulus) and the sum is over all particles labeled
by $j$ and with mass $m_j$ that are inside $S$.  We find that
$A_2$ for the inner $2R_d$ of the disc reaches $0.43$ at $t=6.7\,{\rm
  Gyr}$, decreases to $0.36$ by $t= 9.2\,{\rm Gyr}$, presumably
because the bar has buckled, and then increases to $0.47$ by the
present epoch.  On the other hand, $A_2$ for the entire disc increases
to $0.27$, decreases to $0.23$, and then increases to $0.28$ for the
same epochs.  Note that the inner $2R_d$ of the disc contains $60\%$
of the mass.  Thus, the fact that $A_{2,2R_d}/A_{2,{\rm tot}}\simeq
  0.6$ implies that most of the bar mass resides within the inner
  $2R_d$.

The bars in \citet{YurinSpringelStellarDisks} seem to be stronger then
then ones in our simulations --- they find $A_2\simeq 0.6$ but use a
non-standard formula for $A_2$.  Moreover, their bars appear to extend
across most of the disc.  In terms of disc dynamics, the main
difference between our simulations is the fact that we use a
three-integral DF for the disc whereas they use a two-integral DF.  In
the latter, the velocity dispersion in the radial and vertical
directions are the same.  Thus, the radial dispersion, which fixes the
Toomre $Q$ parameter, also determines the thickness of the disc.  We
note that their initial discs are a factor of two or three thicker
than ours.  We speculate that the bars that develop in these thick
discs are less susceptible to buckling and therefore able to grow
stronger and longer.  These ideas will be investigated in more detail
in a future publication.

\subsection{Kicked-up Stars and Disc Heating}

The outer part of the disc suffers considerable disruption and warping
presumably through its interaction with the main halo or substructure.  The right-most
panel of the $a=0.55$ snapshot in Fig. \ref{fig:cosmo_density_panels}, for example, shows a classic
integral-sign warp.  The other snapshots show that a significant
number of disc particles have orbits that now take them to high
galactic latitudes.

\begin{figure} \includegraphics[width=0.5\textwidth]{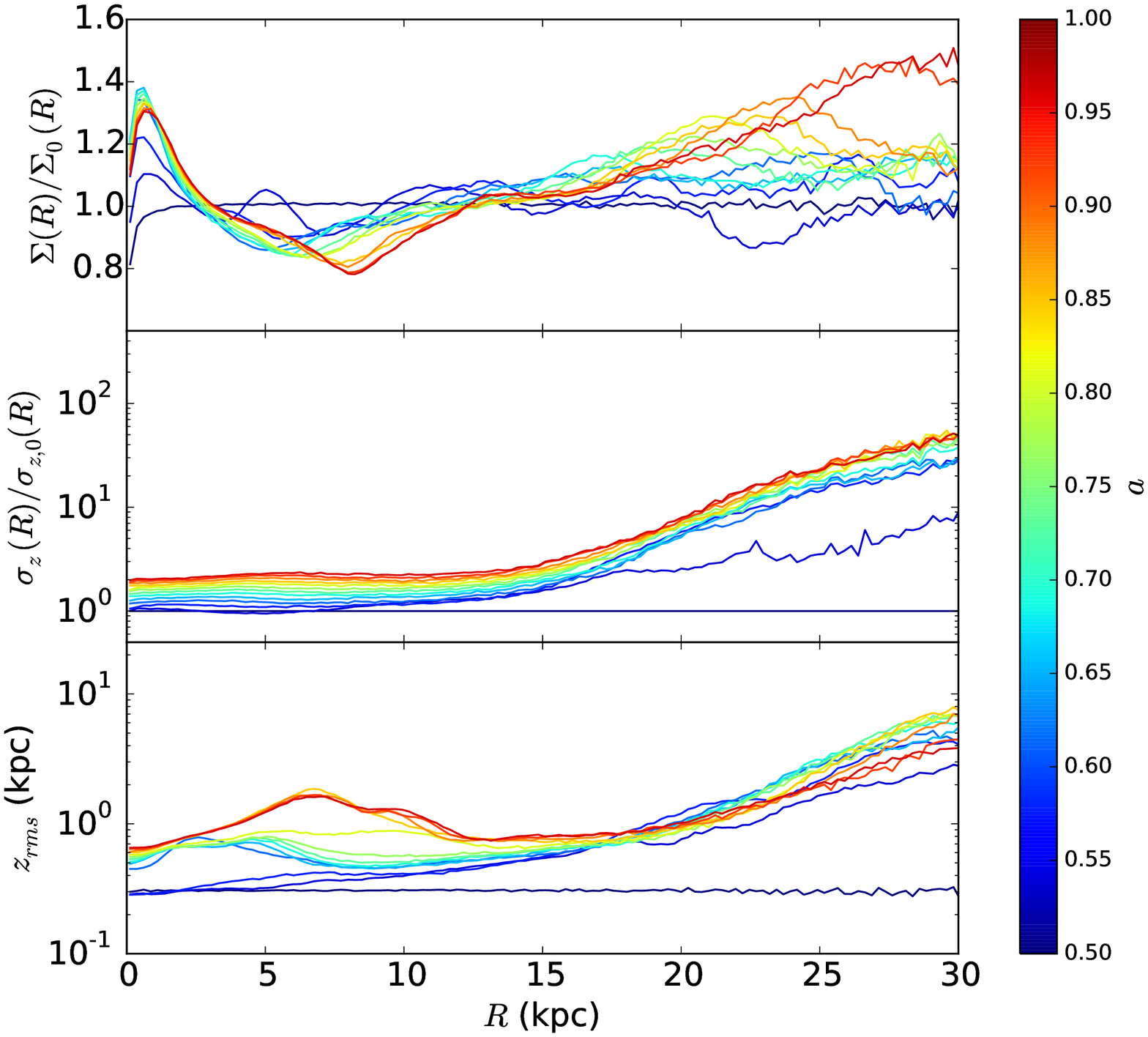} 
  \caption{Surface density, vertical velocity dispersion, and scale
    height profiles of the live disc for 10 snapshots equally spaced
    in scale factor $a$ between $a=0.5$ ($z=1$) and $a=1$ (present
    epoch).  Panels are the same as in
    Fig. \ref{fig:flattened_surface_densities}.}
\label{fig:surface_densities} 
\end{figure}

The impressions one has from the density projections are borne out in
Fig.\,\ref{fig:surface_densities} where we show the surface density
and scale height profiles at different times.  Bar formation
redistributes mass in the disc leaving a deficit (relative to the
initial exponential disk) between $5$ and $15\,{\rm kpc}$.  The disc
becomes thicker and its vertical velocity dispersion increases though
a combination of disc-halo interactions and the effects of the bar and
spiral structure \citep{gauthier2006, dubinski2008, kazantzidis2008}.

A striking feature of the simulations are the streams of disc stars
well above the disc plane.  These stars may represent an example of a
kicked-up disc, which has been seen in other N-body simulations
\citep{PurcellHeatedDisk, McCarthyHeatedDisk} and invoked to explain
kinematic and spectroscopic observations of M31
\citep{DormanKickedUpDisk} and the Monoceros Ring \citep[e.g.][]{monoceros_disc,ibata_et_al_2003}. The idea is that interactions between the
disc and both satellite galaxies and halo substructure liberate stars
from the disc, launching them to regions of the galaxy normally
associated with the stellar halo.  Our live disc simulation
corroborates this hypothesis and is in broad agreement with previous
numerical work. 

Finally, we note that the $a=1$ panel of Fig. \ref{fig:cosmo_density_panels}
shows a relatively thin, stream-like structure, above the disc which
is qualitatively similar to the Anti-centre Stream \citep[ACS,][]{acs_disc}. While the ACS
is believed to be due to the disruption of a globular cluster \citep[e.g.][]{acs_disc}, Fig. \ref{fig:cosmo_density_panels} suggests that perturbations to the disc can create similar features. Intriguingly, \citep{de_boer_et_al_2017} recently 
found that the ACS is rotating in the same sense as the Milky Way disc.


\section{Halo Substructure in the Presence of a Disc}

In this section, we consider the effect of a disc on a halo's
structural properties such as its spherically-averaged density
profile, its shape, and its subhalo population.  An examination of the
DMO simulation shows that the halo we have selected builds up through a
series of mergers and accretion events, but that by $z=1$ it has
settled into a relatively relaxed state with an NFW profile that
evolves very little between $z=1$ and $z=0$ within the inner 100 kpc.  Our sequence of
simulations, (MN, FO, RD, and LD) allow us to tease out the effects of
different disc insertion methods.  The MN simulation, for example,
pins the centre of the disc to the minimum of the halo potential,
whereas the other simulations dynamically evolve the position and
velocity of the disc potential via Newtonian mechanics.  The MN and FO
simulations both assume that the orientation of the disc potential
during the growth phase is fixed whereas RD and LD solve for the
orientation using rigid body dynamics.
 
\subsection{Global Properties of the Halo}

\begin{figure} \includegraphics[width=0.5\textwidth]{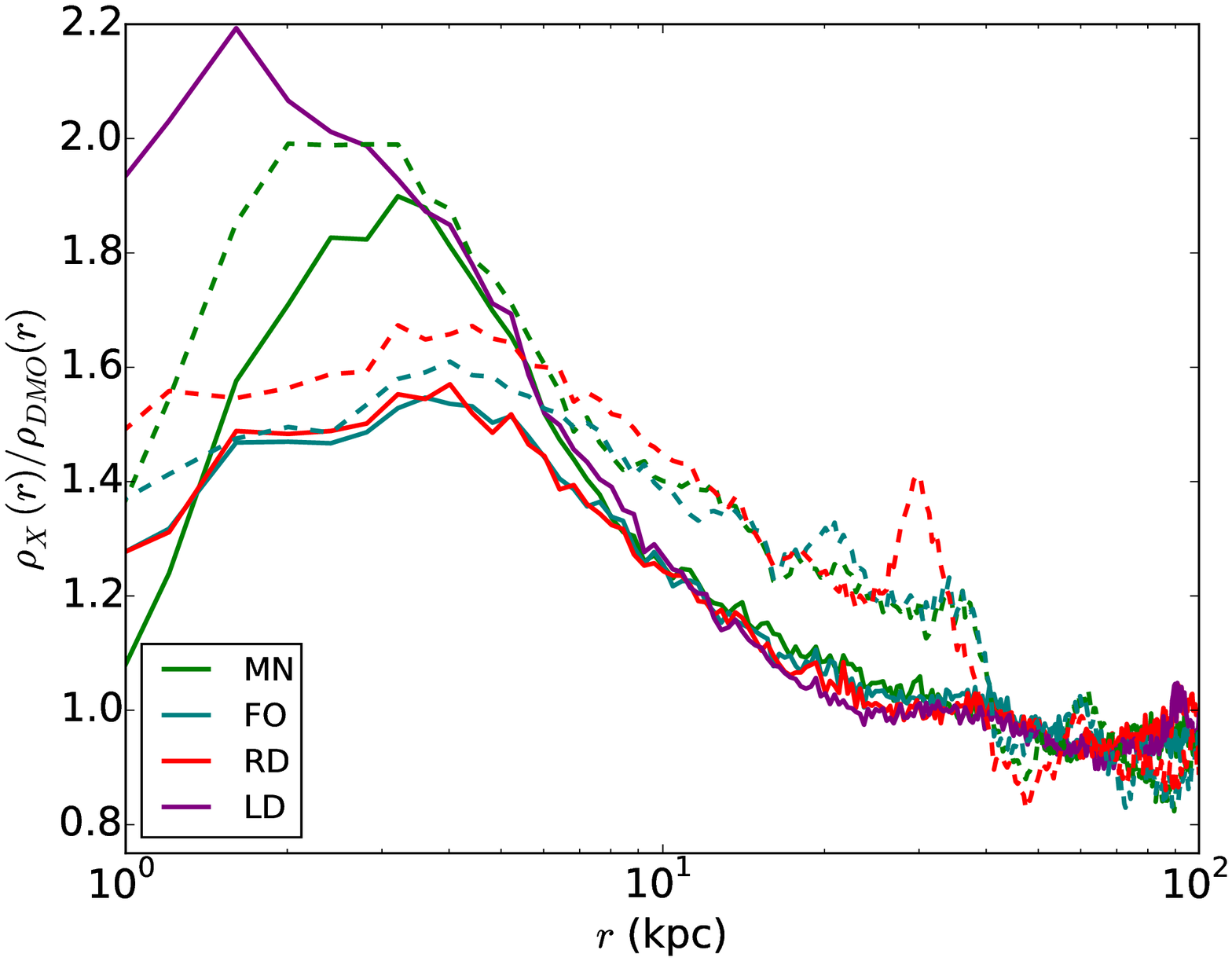} 
  \caption{The ratio of halo density to the DMO simulation for MN
    (green), FO (teal), RD (red), and LD (purple) at $z=1$ (dashed)
    and $z=0$ (solid). The presence of the disc significantly increases the central concentration of the halo.}
\label{fig:halo_ratios}
\end{figure}

In Fig.\,\ref{fig:halo_ratios} we show the ratio of the
spherically-averaged density profile in the four disc runs to that
from the DMO run.  At $z=1$ the haloes in the FO, RD, and MN runs show
evidence for adiabatic contraction with the density in the inner $\sim
30\,{\rm kpc}$ increasing by a factor of $1.2-2.1$.  The effect is
strongest in the MN simulation, which is to be expected since the halo
in that case always sees the disc potential at the minimum of its
potential.  Of course, this prescription is unphysical.  In general,
and especially during a major merger, the disc and halo potential
minimum will not necessarily coincide.

Between $z=1$ and $z=0$, the mass of the disc is constant.  Adiabatic
contraction ceases but the halo still responds to the time-varying
disc potential.  Interestingly, at intermediate radii (between
$\sim 10-40\,{\rm kpc}$) the density profile of the halo settles back
to a state close to that found in the DMO run.  Perhaps most striking
is the fact that the halo in the LD run becomes more centrally
concentrated than the halo in any of the other cases.  One possible explanation
is that dynamical friction from the disc drags dark matter subhaloes
toward the centre of the halo where they are tidally disrupted.

\begin{figure} 
\centering 
\includegraphics[width=0.5\textwidth]{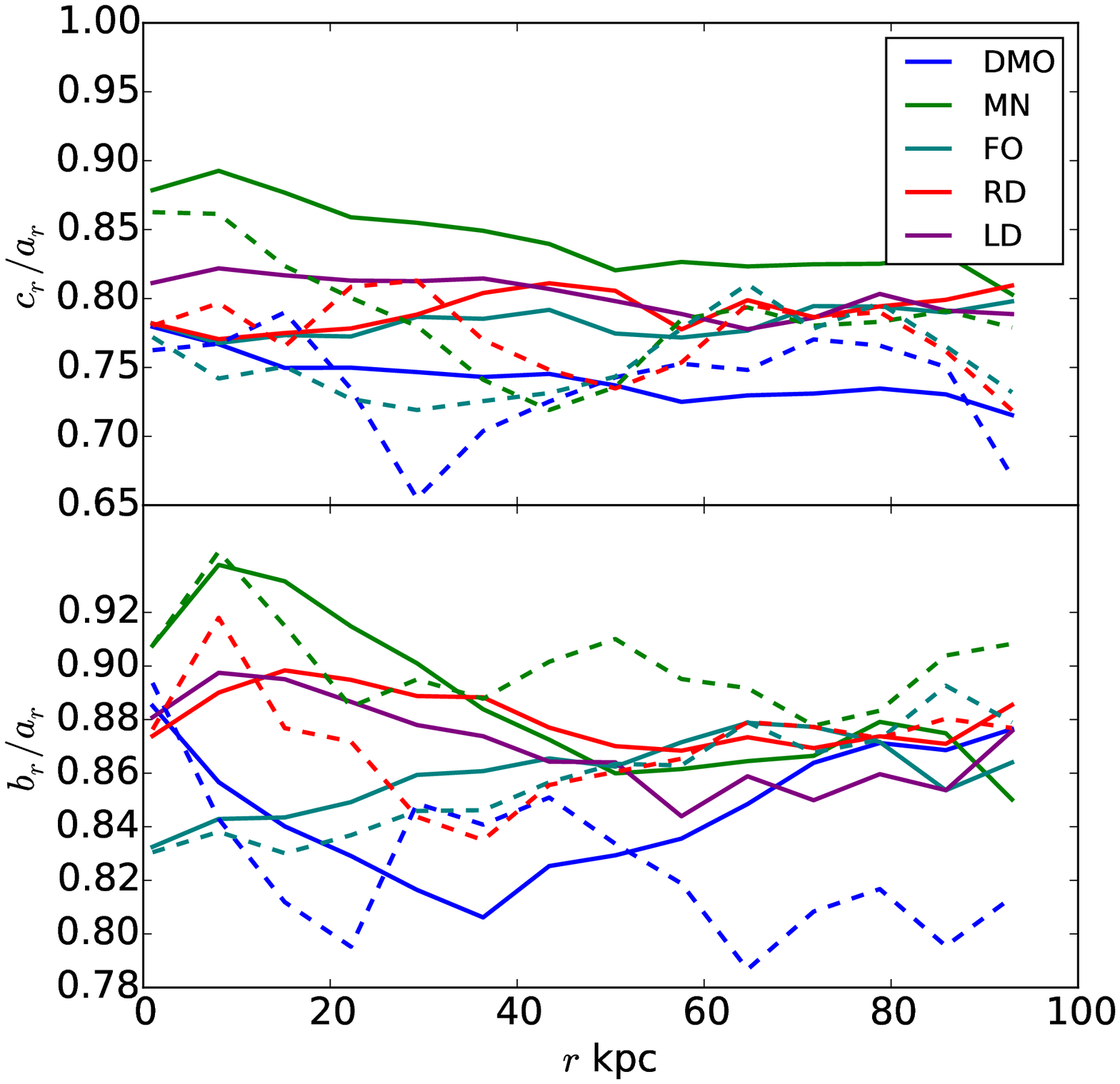} 
\caption{Axis ratios as a function of radius.  Shown are the
  minor-to-major axis ratio (top panel) and the intermediate-to-major
  axis ratio (bottom panel) at $z=1$ (dashed curves) and $z=0$ (solid
  curves).  Blue corresponds to DMO, green to MN, teal to FO, red to RD, and
  purple to LD.}
\label{fig:axis_ratios}
\end{figure}

In Fig.\,\ref{fig:axis_ratios} we show the minor-to-major ($c_r/a_r$)
and intermediate-to-major ($b_r/a_r$) axis ratios as a function of
radius for both the $z=1$ and $z=0$ snapshots.  The axis ratios are
calculated by diagonalizing the moment of inertia tensor in
linearly-spaced radial shells.  The DMO halo is triaxial with
$c_r/a_r\simeq 0.75$ and $b_r/a_r\simeq 0.85$.  Note that the axis
ratio profiles are smoother at $z=0$ than at $z=1$, which supports the
observation that the halo has settled into a more relaxed state over
the past $7$ or so billion years.  In general, discs tend to make
halos more spherical, a result that has been known for some time from
both collisionless and hydrodynamical simulations
\citep[e.g.][]{dubinski1994_ApJ431_617,Zemp2012}.

Evidently, the MN halo is rounder, especially in the inner part, the
the FO halo.  Recall that the main difference between these two cases
is that the MN disc is pinned to the potential minimum of the halo.
It is perhaps not surprising then that, as with adiabatic contraction,
it has a stronger effect on the halo's shape.  
We also note that the axis ratio profiles for the RD and LD simulations
are fairly similar.  

\subsection{Subhalo Populations}

We now turn our attention to halo substructure.  We identify subhaloes
and determine their positions and masses using \textsc{rockstar}
\citep{rockstar}, which employs a friends-of-friends algorithm in six
phase space dimensions.  We consider only those subhaloes with mass
$m_s$ between $m_{\rm min} = 10^{7.5}\,M_\odot$ and $m_{\rm max} =
10^{10.5}\,M_\odot$.  Subhaloes at the lower end of this range are
marginally resolved with $\sim100$ particles, above which the subhalo
mass can be trusted \citep[e.g.][]{onions_et_al_2012}, while those at the upper end contain $\sim 3\%$ of
the halo's virial mass.

\begin{figure} 
\centering 
{\includegraphics[width=0.5\textwidth]{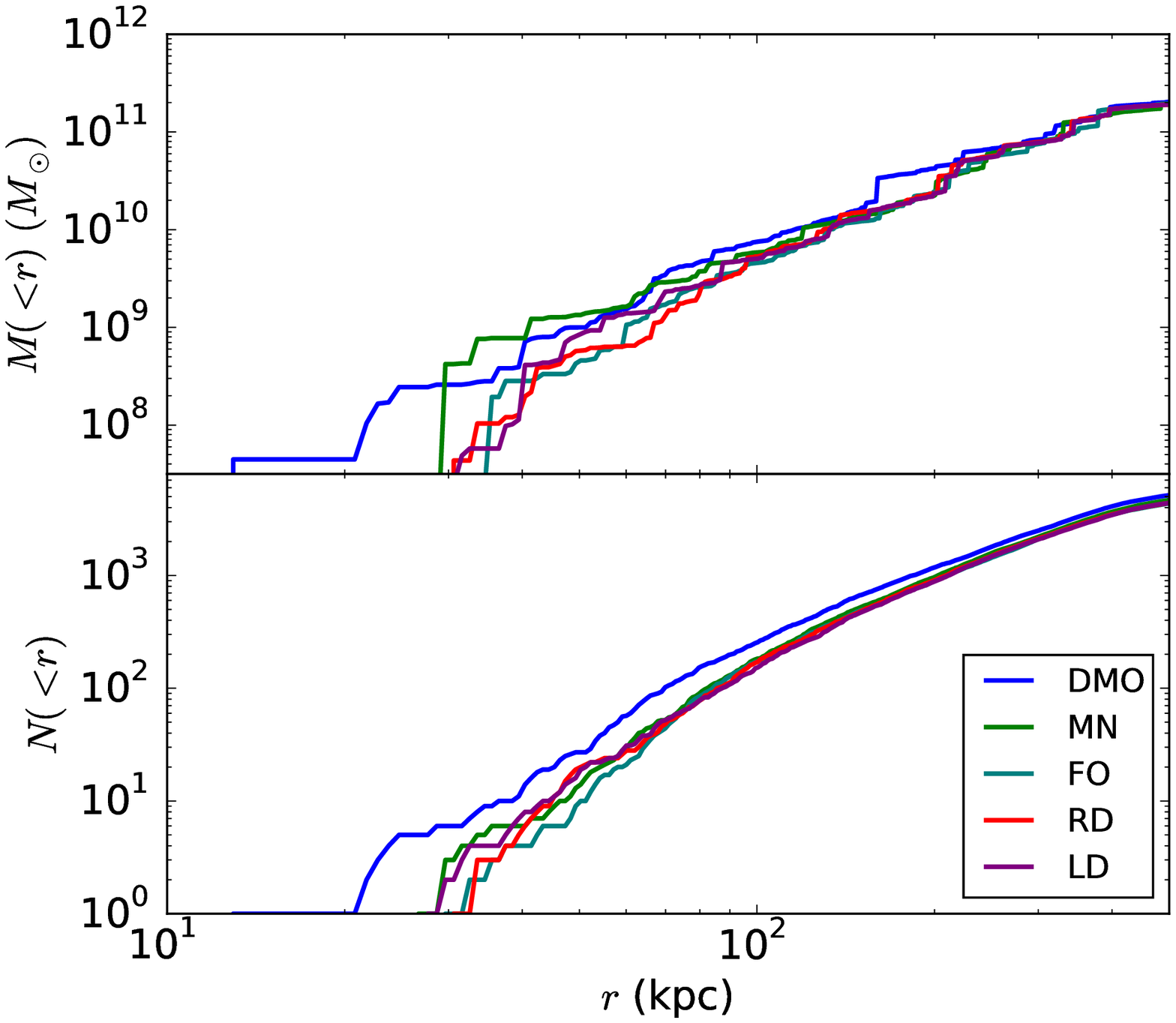}}
\caption{Cumulative mass in subhaloes inside a radius $r$ (upper panel)
  and cumulative number of subhaloes (lower panel).  We consider only
  subhaloes within $500\,{\rm kpc}$ of the halo centre and with a mass
  above $10^{7.5} M_\odot$.  The curves are blue (DMO), green (MN),
  teal (FO), red (RD), and purple (LD).}
\label{fig:cumulative_distributions}
\end{figure}

\begin{figure} 
\centering 
{\includegraphics[width=0.5\textwidth]{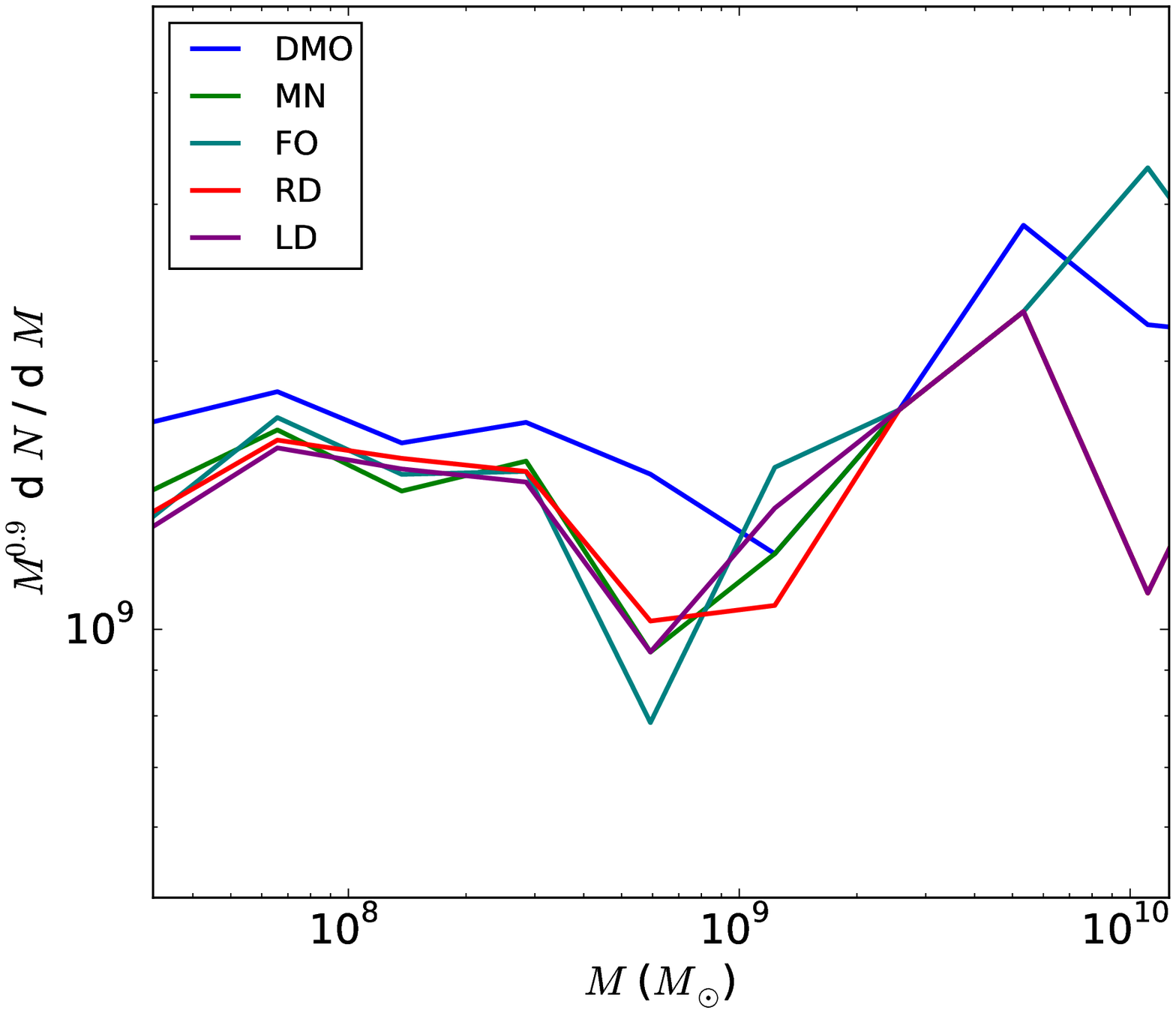}}
\caption{ Differential mass distribution
  multiplied by $M^{0.9}$ for subhaloes above $10^{7.5} M_\odot$. We
  make an outer radius cut at 500 kpc. The curves are blue
  (DMO), green (MN), teal (FO), red (RD), and purple (LD). }
\label{fig:differential_mass_distributions}
\end{figure}

In Fig.\,\ref{fig:cumulative_distributions} we show the cumulative
mass in subhaloes as a function of Galactocentric radius:

\begin{equation}
M_s\left (<r\right )  = \int_0^r dr \int_{m_{\rm min}}^{m_{\rm
    max}} dm_s\,m_s\,\frac{d^2 N}{dm_s\,dr}
\end{equation}

\noindent We also show the cumulative number of subhaloes.  In
general, the presence of a disc depletes substructure inside about
$30\,{\rm kpc}$ but leaves the outer substructures unaffected.

We next consider the differential mass distribution as a function of
subhalo mass.  Recall that for a pure dark matter halo,
$dN/d\ln(m_s)\propto m_s^{-p}$ where $p\simeq 0.9$
\citep[e.g.][]{gao2004}.  In
Fig. \ref{fig:differential_mass_distributions}, we therefore show the
quantity $m^{0.9}\,dN/d\ln(m_s)$ in order to enhance the differences
between the different disc runs.  We see that the halo population
between $m_s \simeq m_{\rm min}$ and $m_s\simeq 10^9\,M_\odot$ is
depleted, but only by about $20-30\,\%$.  Taken together,
Fig.\,\ref{fig:cumulative_distributions} and
Fig. \ref{fig:differential_mass_distributions} imply that the main
depletion of the subhaloes occurs within the inner regions of the
parent halo, in agreement with \citet{DOhngiaSubstructureDepletion,Sawala2017,GKSubhaloDepletion17}.  That said, the depletion of subhaloes seems rather
insensitive to the disc insertion method, although we caution that only a single halo was used. Our results, being mainly in agreement with previous work, should still be viewed with caution due the consideration of a single host halo.

\subsection{Case Study: A Sagittarius-like Dwarf}

\begin{figure*}
\includegraphics[width=\textwidth]{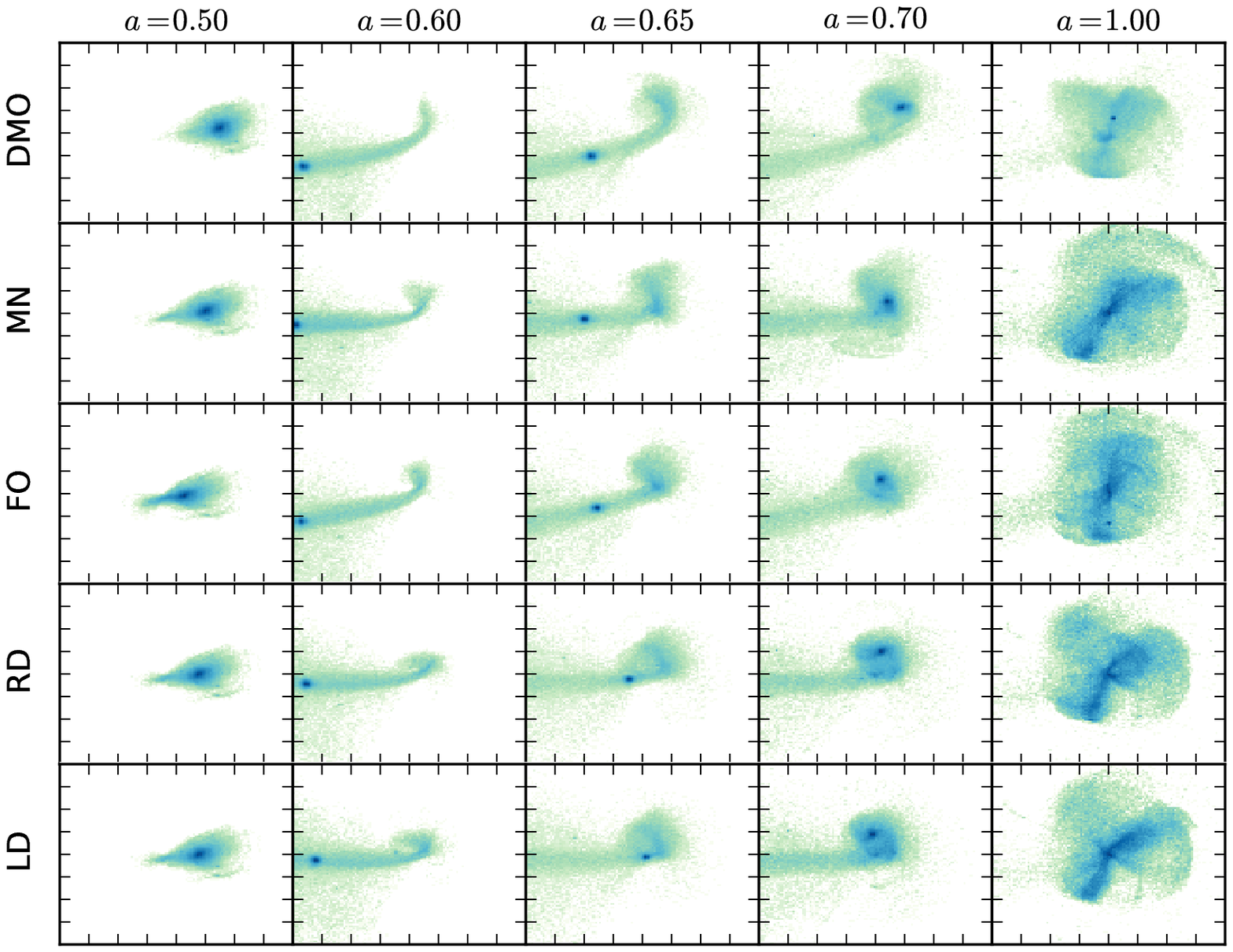}
\caption{X-Y projections for a selected Sagittarius-like dwarf galaxy. The
  rows from top to bottom are no disc, a fixed Miyamoto-Nagai disc, a
  rigid disc, and a live disc. The scale factors in columns from left
  to right are 0.5, 0.55, 0.6, 0.65, and 0.7. The frame edges are
  $295\,{\rm kpc}$ on each side.}\label{fig:streams}
\end{figure*}

Observations of the Milky Way's dwarf galaxies and associated tidal
streams provide a potentially powerful probe of the Galactic potential
and thus the Galaxy's dark halo.  One of the best-studied examples is
the Sagittarius dwarf \citet{ibatadiscovery}.  Fortunately, our
simulation has a satellite galaxy with similar properties, namely, a
dark matter mass of $1.8 \times 10^{10} \, M_\odot$ at
$z=1.0$ and an orbit that takes it close to the disc.  We identify
this object in the five simulations using the \textsc{rockstar}
halo catalogues.  We then gather a list of IDs for all the
bound particles at an early time before the dwarf is disrupted and
follow these same particles in later snapshots using a binary
search tree look-up scheme.  In Fig.\,\ref{fig:streams} we show the
evolution of this subhalo between redshift $z_l=1$ and $z=0$. The
first row shows the baseline evolution in our
DMO simulation.  The dwarf develops leading and trailing tidal tails
during the first few billion years.  By the present epoch, the tidal
debris has dispersed throughout the halo.

The next four rows show the same satellite in our disc simulations.
Perhaps the most noticeable result is that there are stronger features
in the tidal debris at the present epoch once a disc is included.  The
detailed morphology of the tidal debris is certainly different from
one disc simulation to the next.  By eye, debris in MN and FO look
somewhat similar as does the debris in RD and LD. Perhaps the most
noticeable result is that the tidal debris extends to larger galactocentric
radii when a disc is included. The detailed morphology of the tidal debris
clearly depends on the disc insertion method. By eye, tidal debris appears
more isotropic in MN and FO than in RD and LD. The implication is that fixed
potentials are more efficient at disrupting massive satellites than a 
potential which can respond to the satellite's presence.  However, we have
only a single example of massive satellite disruption, and we caution that 
more examples of this behaviour are needed to test this hypothesis.


\section{Conclusions}

Simulations in which a stellar disc is inserted ``by hand" into a
cosmological N-body halo provide a compromise between simulations of
isolated disc-halo systems and cosmological simulations that include
gasdynamics and star formation.  Our method builds on the scheme
used by \citet{BerentzenShlosmanStellarDisks,DeBuhrStellarDisks} and refined by \citet{YurinSpringelStellarDisks}.  The basic idea is to introduce, at a redshift $z_g$,
a rigid disc with zero mass into a halo within a cosmological zoom-in
simulation.  Between $z_g$ and $z_l$ the disc is treated as an external
potential with a mass and size that increase adiabatically to their
present day values.  At $z_l$, the rigid disc is replaced by an N-body
one and the simulation proceeds to the present epoch with live disc
and halo particles.

Our method improves upon previous ones in two important ways.  First,
during the growth phase ($z_g > z > z_l$) the position and orientation
of the disc evolve according to the standard equations of rigid-body
dynamics.  Thus, the disc in our scheme receives its linear and angular
momentum with the halo in a self-consistent fashion and is therefore
able to move, precess, and nutate due to torques from the halo.  While
previous methods introduced aspects of rigid-body dynamics during the
growth phase none appear to have implemented the full dynamical
equations have done here \citep{DOhngiaSubstructureDepletion, DeBuhrStellarDisks, YurinSpringelStellarDisks}.

Our sequence MN, FO, RD, and LD of simulations highlights where the
details of the disc insertion scheme are important and where they are
not.  For example, schemes in which the disc tracks the minimum of the
halo potential tend to overestimate the effects of adiabatic
contraction.  On the other hand, the effect of the depletion of halo
substructure seems to be rather insensitive to the details of how the
disc is introduced into the simulation.

Disc insertion schemes such as the one introduced in this paper,
provide an attractive arena for
studies of galactic dynamics. In particular, they allow one to study the interaction between a stellar disc and a realistic dark halo with computationally inexpensive simulations while maintaining some level of control over the structural parameters of the disc.  We fully intend to leverage these advantages in future work.

\section*{Acknowledgements} {LMW and JB are supported by a Discovery
  Grant with the Natural Sciences and Engineering Research Council of
  Canada. The research leading to these results has received funding
  from the European Research Council under the European Union's
  Seventh Framework Programme (FP/2007-2013) / ERC Grant Agreement
  n. 308024. DE acknowledges financial support from the ERC.}




\bibliographystyle{mnras}
\bibliography{bibliography.bib} 

\begin{thebibliography}{}
\makeatletter
\relax
\def\mn@urlcharsother{\let\do\@makeother \do\$\do\&\do\#\do\^\do\_\do\%\do\~}
\def\mn@doi{\begingroup\mn@urlcharsother \@ifnextchar [ {\mn@doi@}
  {\mn@doi@[]}}
\def\mn@doi@[#1]#2{\def\@tempa{#1}\ifx\@tempa\@empty \href
  {http://dx.doi.org/#2} {doi:#2}\else \href {http://dx.doi.org/#2} {#1}\fi
  \endgroup}
\def\mn@eprint#1#2{\mn@eprint@#1:#2::\@nil}
\def\mn@eprint@arXiv#1{\href {http://arxiv.org/abs/#1} {{\tt arXiv:#1}}}
\def\mn@eprint@dblp#1{\href {http://dblp.uni-trier.de/rec/bibtex/#1.xml}
  {dblp:#1}}
\def\mn@eprint@#1:#2:#3:#4\@nil{\def\@tempa {#1}\def\@tempb {#2}\def\@tempc
  {#3}\ifx \@tempc \@empty \let \@tempc \@tempb \let \@tempb \@tempa \fi \ifx
  \@tempb \@empty \def\@tempb {arXiv}\fi \@ifundefined
  {mn@eprint@\@tempb}{\@tempb:\@tempc}{\expandafter \expandafter \csname
  mn@eprint@\@tempb\endcsname \expandafter{\@tempc}}}

\bibitem[\protect\citeauthoryear{{Behroozi}, {Wechsler}  \& {Wu}}{{Behroozi}
  et~al.}{2013}]{rockstar}
{Behroozi} P.~S.,  {Wechsler} R.~H.,   {Wu} H.-Y.,  2013, \mn@doi [\apj]
  {10.1088/0004-637X/762/2/109}, \href
  {http://adsabs.harvard.edu/abs/2013ApJ...762..109B} {762, 109}

\bibitem[\protect\citeauthoryear{{Berentzen} \& {Shlosman}}{{Berentzen} \&
  {Shlosman}}{2006}]{BerentzenShlosmanStellarDisks}
{Berentzen} I.,  {Shlosman} I.,  2006, \mn@doi [\apj] {10.1086/506016}, \href
  {http://adsabs.harvard.edu/abs/2006ApJ...648..807B} {648, 807}

\bibitem[\protect\citeauthoryear{{Binney} \& {Tremaine}}{{Binney} \&
  {Tremaine}}{2008}]{BT}
{Binney} J.,  {Tremaine} S.,  2008, {Galactic Dynamics: Second Edition}.
Princeton University Press

\bibitem[\protect\citeauthoryear{{D'Onghia}, {Springel}, {Hernquist}  \&
  {Keres}}{{D'Onghia} et~al.}{2010}]{DOhngiaSubstructureDepletion}
{D'Onghia} E.,  {Springel} V.,  {Hernquist} L.,   {Keres} D.,  2010, \mn@doi
  [\apj] {10.1088/0004-637X/709/2/1138}, \href
  {http://adsabs.harvard.edu/abs/2010ApJ...709.1138D} {709, 1138}

\bibitem[\protect\citeauthoryear{{DeBuhr}, {Ma}  \& {White}}{{DeBuhr}
  et~al.}{2012}]{DeBuhrStellarDisks}
{DeBuhr} J.,  {Ma} C.-P.,   {White} S.~D.~M.,  2012, \mn@doi [\mnras]
  {10.1111/j.1365-2966.2012.21910.x}, \href
  {http://adsabs.harvard.edu/abs/2012MNRAS.426..983D} {426, 983}

\bibitem[\protect\citeauthoryear{{Dorman} et~al.,}{{Dorman}
  et~al.}{2013}]{DormanKickedUpDisk}
{Dorman} C.~E.,  et~al., 2013, \mn@doi [\apj] {10.1088/0004-637X/779/2/103},
  \href {http://adsabs.harvard.edu/abs/2013ApJ...779..103D} {779, 103}

\bibitem[\protect\citeauthoryear{{Dubinski}}{{Dubinski}}{1994}]{dubinski1994_ApJ431_617}
{Dubinski} J.,  1994, \mn@doi [\apj] {10.1086/174512}, \href
  {http://adsabs.harvard.edu/abs/1994ApJ...431..617D} {431, 617}

\bibitem[\protect\citeauthoryear{{Dubinski} \& {Kuijken}}{{Dubinski} \&
  {Kuijken}}{1995}]{DubinskiKuijkenRigidDisks}
{Dubinski} J.,  {Kuijken} K.,  1995, \mn@doi [\apj] {10.1086/175456}, \href
  {http://adsabs.harvard.edu/abs/1995ApJ...442..492D} {442, 492}

\bibitem[\protect\citeauthoryear{{Dubinski}, {Gauthier}, {Widrow}  \&
  {Nickerson}}{{Dubinski} et~al.}{2008}]{dubinski2008}
{Dubinski} J.,  {Gauthier} J.-R.,  {Widrow} L.,   {Nickerson} S.,  2008, in
  {Funes} J.~G.,  {Corsini} E.~M.,  eds,  Astronomical Society of the Pacific
  Conference Series Vol. 396, Formation and Evolution of Galaxy Disks. p.~321
  (\mn@eprint {arXiv} {0802.3997})

\bibitem[\protect\citeauthoryear{{Font}, {Navarro}, {Stadel}  \&
  {Quinn}}{{Font} et~al.}{2001}]{Font2001}
{Font} A.~S.,  {Navarro} J.~F.,  {Stadel} J.,   {Quinn} T.,  2001, \mn@doi
  [\apjl] {10.1086/338479}, \href
  {http://adsabs.harvard.edu/abs/2001ApJ...563L...1F} {563, L1}

\bibitem[\protect\citeauthoryear{{Gao}, {White}, {Jenkins}, {Stoehr}  \&
  {Springel}}{{Gao} et~al.}{2004}]{gao2004}
{Gao} L.,  {White} S.~D.~M.,  {Jenkins} A.,  {Stoehr} F.,   {Springel} V.,
  2004, \mn@doi [\mnras] {10.1111/j.1365-2966.2004.08360.x}, \href
  {http://adsabs.harvard.edu/abs/2004MNRAS.355..819G} {355, 819}

\bibitem[\protect\citeauthoryear{{Garrison-Kimmel} et~al.,}{{Garrison-Kimmel}
  et~al.}{2017}]{GKSubhaloDepletion17}
{Garrison-Kimmel} S.,  et~al., 2017, preprint, \href
  {http://adsabs.harvard.edu/abs/2017arXiv170103792G} {} (\mn@eprint {arXiv}
  {1701.03792})

\bibitem[\protect\citeauthoryear{{Gauthier}, {Dubinski}  \&
  {Widrow}}{{Gauthier} et~al.}{2006}]{gauthier2006}
{Gauthier} J.-R.,  {Dubinski} J.,   {Widrow} L.~M.,  2006, \mn@doi [\apj]
  {10.1086/508860}, \href {http://adsabs.harvard.edu/abs/2006ApJ...653.1180G}
  {653, 1180}

\bibitem[\protect\citeauthoryear{{G{\'o}mez}, {White}, {Grand}, {Marinacci},
  {Springel}  \& {Pakmor}}{{G{\'o}mez} et~al.}{2016}]{gomezwarps}
{G{\'o}mez} F.~A.,  {White} S.~D.~M.,  {Grand} R.~J.~J.,  {Marinacci} F.,
  {Springel} V.,   {Pakmor} R.,  2016, preprint, \href
  {http://adsabs.harvard.edu/abs/2016arXiv160606295G} {} (\mn@eprint {arXiv}
  {1606.06295})

\bibitem[\protect\citeauthoryear{{Grillmair}}{{Grillmair}}{2006}]{acs_disc}
{Grillmair} C.~J.,  2006, \mn@doi [\apjl] {10.1086/509255}, \href
  {http://adsabs.harvard.edu/abs/2006ApJ...651L..29G} {651, L29}

\bibitem[\protect\citeauthoryear{{Hahn} \& {Abel}}{{Hahn} \&
  {Abel}}{2013}]{music}
{Hahn} O.,  {Abel} T.,  2013, {MUSIC: MUlti-Scale Initial Conditions},
  Astrophysics Source Code Library (\mn@eprint {ascl} {1311.011})

\bibitem[\protect\citeauthoryear{{Hu} \& {Sijacki}}{{Hu} \&
  {Sijacki}}{2016}]{hu_sijacki_2016}
{Hu} S.,  {Sijacki} D.,  2016, \mn@doi [\mnras] {10.1093/mnras/stw1463}, \href
  {http://adsabs.harvard.edu/abs/2016MNRAS.461.2789H} {461, 2789}

\bibitem[\protect\citeauthoryear{{Ibata}, {Gilmore}  \& {Irwin}}{{Ibata}
  et~al.}{1994}]{ibatadiscovery}
{Ibata} R.~A.,  {Gilmore} G.,   {Irwin} M.~J.,  1994, \mn@doi [\nat]
  {10.1038/370194a0}, \href {http://adsabs.harvard.edu/abs/1994Natur.370..194I}
  {370, 194}

\bibitem[\protect\citeauthoryear{{Ibata}, {Irwin}, {Lewis}, {Ferguson}  \&
  {Tanvir}}{{Ibata} et~al.}{2003}]{ibata_et_al_2003}
{Ibata} R.~A.,  {Irwin} M.~J.,  {Lewis} G.~F.,  {Ferguson} A.~M.~N.,   {Tanvir}
  N.,  2003, \mn@doi [\mnras] {10.1046/j.1365-8711.2003.06545.x}, \href
  {http://adsabs.harvard.edu/abs/2003MNRAS.340L..21I} {340, L21}

\bibitem[\protect\citeauthoryear{{Katz}, {Quinn}, {Bertschinger}  \&
  {Gelb}}{{Katz} et~al.}{1994}]{KatzQuasarZoom}
{Katz} N.,  {Quinn} T.,  {Bertschinger} E.,   {Gelb} J.~M.,  1994, \mn@doi
  [\mnras] {10.1093/mnras/270.1.L71}, \href
  {http://adsabs.harvard.edu/abs/1994MNRAS.270L..71K} {270, L71}

\bibitem[\protect\citeauthoryear{{Katz}, {Weinberg}  \& {Hernquist}}{{Katz}
  et~al.}{1996}]{katz1996feedback}
{Katz} N.,  {Weinberg} D.~H.,   {Hernquist} L.,  1996, \mn@doi [\apjs]
  {10.1086/192305}, \href {http://adsabs.harvard.edu/abs/1996ApJS..105...19K}
  {105, 19}

\bibitem[\protect\citeauthoryear{{Kazantzidis}, {Bullock}, {Zentner},
  {Kravtsov}  \& {Moustakas}}{{Kazantzidis} et~al.}{2008}]{kazantzidis2008}
{Kazantzidis} S.,  {Bullock} J.~S.,  {Zentner} A.~R.,  {Kravtsov} A.~V.,
  {Moustakas} L.~A.,  2008, \mn@doi [\apj] {10.1086/591958}, \href
  {http://adsabs.harvard.edu/abs/2008ApJ...688..254K} {688, 254}

\bibitem[\protect\citeauthoryear{{Klypin}, {Kravtsov}, {Valenzuela}  \&
  {Prada}}{{Klypin} et~al.}{1999}]{Klypin1999}
{Klypin} A.,  {Kravtsov} A.~V.,  {Valenzuela} O.,   {Prada} F.,  1999, \mn@doi
  [\apj] {10.1086/307643}, \href
  {http://adsabs.harvard.edu/abs/1999ApJ...522...82K} {522, 82}

\bibitem[\protect\citeauthoryear{{Kuijken} \& {Gilmore}}{{Kuijken} \&
  {Gilmore}}{1989}]{KGGalactICSReference}
{Kuijken} K.,  {Gilmore} G.,  1989, \mn@doi [\mnras] {10.1093/mnras/239.2.571},
  \href {http://adsabs.harvard.edu/abs/1989MNRAS.239..571K} {239, 571}

\bibitem[\protect\citeauthoryear{{Laporte}, {G{\'o}mez}, {Besla}, {Johnston}
  \& {Garavito-Camargo}}{{Laporte} et~al.}{2016}]{laporte_et_al_2016}
{Laporte} C.~F.~P.,  {G{\'o}mez} F.~A.,  {Besla} G.,  {Johnston} K.~V.,
  {Garavito-Camargo} N.,  2016, preprint, \href
  {http://adsabs.harvard.edu/abs/2016arXiv160804743L} {} (\mn@eprint {arXiv}
  {1608.04743})

\bibitem[\protect\citeauthoryear{{McCarthy}, {Font}, {Crain}, {Deason},
  {Schaye}  \& {Theuns}}{{McCarthy} et~al.}{2012}]{McCarthyHeatedDisk}
{McCarthy} I.~G.,  {Font} A.~S.,  {Crain} R.~A.,  {Deason} A.~J.,  {Schaye} J.,
    {Theuns} T.,  2012, \mn@doi [\mnras] {10.1111/j.1365-2966.2011.20189.x},
  \href {http://adsabs.harvard.edu/abs/2012MNRAS.420.2245M} {420, 2245}

\bibitem[\protect\citeauthoryear{{Miyamoto} \& {Nagai}}{{Miyamoto} \&
  {Nagai}}{1975}]{MiyamotoNagai}
{Miyamoto} M.,  {Nagai} R.,  1975, \pasj, \href
  {http://adsabs.harvard.edu/abs/1975PASJ...27..533M} {27, 533}

\bibitem[\protect\citeauthoryear{{Moore}, {Ghigna}, {Governato}, {Lake},
  {Quinn}, {Stadel}  \& {Tozzi}}{{Moore} et~al.}{1999}]{mooresubhalos}
{Moore} B.,  {Ghigna} S.,  {Governato} F.,  {Lake} G.,  {Quinn} T.,  {Stadel}
  J.,   {Tozzi} P.,  1999, \mn@doi [\apjl] {10.1086/312287}, \href
  {http://adsabs.harvard.edu/abs/1999ApJ...524L..19M} {524, L19}

\bibitem[\protect\citeauthoryear{{Navarro}, {Frenk}  \& {White}}{{Navarro}
  et~al.}{1994}]{NavarroWhiteZoom}
{Navarro} J.~F.,  {Frenk} C.~S.,   {White} S.~D.~M.,  1994, \mn@doi [\mnras]
  {10.1093/mnras/267.1.L1}, \href
  {http://adsabs.harvard.edu/abs/1994MNRAS.267L...1N} {267, L1}

\bibitem[\protect\citeauthoryear{{Navarro}, {Frenk}  \& {White}}{{Navarro}
  et~al.}{1997}]{NFW}
{Navarro} J.~F.,  {Frenk} C.~S.,   {White} S.~D.~M.,  1997, \mn@doi [\apj]
  {10.1086/304888}, \href {http://adsabs.harvard.edu/abs/1997ApJ...490..493N}
  {490, 493}

\bibitem[\protect\citeauthoryear{{Newberg} et~al.,}{{Newberg}
  et~al.}{2002}]{monoceros_disc}
{Newberg} H.~J.,  et~al., 2002, \mn@doi [\apj] {10.1086/338983}, \href
  {http://adsabs.harvard.edu/abs/2002ApJ...569..245N} {569, 245}

\bibitem[\protect\citeauthoryear{{O{\~n}orbe}, {Garrison-Kimmel}, {Maller},
  {Bullock}, {Rocha}  \& {Hahn}}{{O{\~n}orbe} et~al.}{2014}]{onorbe_etal_2014}
{O{\~n}orbe} J.,  {Garrison-Kimmel} S.,  {Maller} A.~H.,  {Bullock} J.~S.,
  {Rocha} M.,   {Hahn} O.,  2014, \mn@doi [\mnras] {10.1093/mnras/stt2020},
  \href {http://adsabs.harvard.edu/abs/2014MNRAS.437.1894O} {437, 1894}

\bibitem[\protect\citeauthoryear{{Onions} et~al.,}{{Onions}
  et~al.}{2012}]{onions_et_al_2012}
{Onions} J.,  et~al., 2012, \mn@doi [\mnras]
  {10.1111/j.1365-2966.2012.20947.x}, \href
  {http://adsabs.harvard.edu/abs/2012MNRAS.423.1200O} {423, 1200}

\bibitem[\protect\citeauthoryear{{Pakmor} \& {Springel}}{{Pakmor} \&
  {Springel}}{2013}]{pakmorMHD}
{Pakmor} R.,  {Springel} V.,  2013, \mn@doi [\mnras] {10.1093/mnras/stt428},
  \href {http://adsabs.harvard.edu/abs/2013MNRAS.432..176P} {432, 176}

\bibitem[\protect\citeauthoryear{{Planck Collaboration} et~al.,}{{Planck
  Collaboration} et~al.}{2014}]{planck_2014}
{Planck Collaboration} et~al., 2014, \mn@doi [\aap]
  {10.1051/0004-6361/201321591}, \href
  {http://adsabs.harvard.edu/abs/2014A26A...571A..16P} {571, A16}

\bibitem[\protect\citeauthoryear{{Power}, {Navarro}, {Jenkins}, {Frenk},
  {White}, {Springel}, {Stadel}  \& {Quinn}}{{Power}
  et~al.}{2003}]{power_et_al_2003}
{Power} C.,  {Navarro} J.~F.,  {Jenkins} A.,  {Frenk} C.~S.,  {White} S.~D.~M.,
   {Springel} V.,  {Stadel} J.,   {Quinn} T.,  2003, \mn@doi [\mnras]
  {10.1046/j.1365-8711.2003.05925.x}, \href
  {http://adsabs.harvard.edu/abs/2003MNRAS.338...14P} {338, 14}

\bibitem[\protect\citeauthoryear{{Purcell}, {Bullock}  \&
  {Kazantzidis}}{{Purcell} et~al.}{2010}]{PurcellHeatedDisk}
{Purcell} C.~W.,  {Bullock} J.~S.,   {Kazantzidis} S.,  2010, \mn@doi [\mnras]
  {10.1111/j.1365-2966.2010.16429.x}, \href
  {http://adsabs.harvard.edu/abs/2010MNRAS.404.1711P} {404, 1711}

\bibitem[\protect\citeauthoryear{{Purcell}, {Bullock}, {Tollerud}, {Rocha}  \&
  {Chakrabarti}}{{Purcell} et~al.}{2011}]{purcell2011}
{Purcell} C.~W.,  {Bullock} J.~S.,  {Tollerud} E.~J.,  {Rocha} M.,
  {Chakrabarti} S.,  2011, \mn@doi [\nat] {10.1038/nature10417}, \href
  {http://adsabs.harvard.edu/abs/2011Natur.477..301P} {477, 301}

\bibitem[\protect\citeauthoryear{{Quinn}, {Katz}, {Stadel}  \& {Lake}}{{Quinn}
  et~al.}{1997}]{QuinnIntegrators}
{Quinn} T.,  {Katz} N.,  {Stadel} J.,   {Lake} G.,  1997, ArXiv Astrophysics
  e-prints, \href {http://adsabs.harvard.edu/abs/1997astro.ph.10043Q} {}

\bibitem[\protect\citeauthoryear{{Ro{\v s}kar}, {Debattista}, {Brooks},
  {Quinn}, {Brook}, {Governato}, {Dalcanton}  \& {Wadsley}}{{Ro{\v s}kar}
  et~al.}{2010}]{RoskarDiskMisalignment}
{Ro{\v s}kar} R.,  {Debattista} V.~P.,  {Brooks} A.~M.,  {Quinn} T.~R.,
  {Brook} C.~B.,  {Governato} F.,  {Dalcanton} J.~J.,   {Wadsley} J.,  2010,
  \mn@doi [\mnras] {10.1111/j.1365-2966.2010.17178.x}, \href
  {http://adsabs.harvard.edu/abs/2010MNRAS.408..783R} {408, 783}

\bibitem[\protect\citeauthoryear{{Sackett}}{{Sackett}}{1997}]{sackett1997}
{Sackett} P.~D.,  1997, \mn@doi [\apj] {10.1086/304223}, \href
  {http://adsabs.harvard.edu/abs/1997ApJ...483..103S} {483, 103}

\bibitem[\protect\citeauthoryear{{Sawala}, {Pihajoki}, {Johansson}, {Frenk},
  {Navarro}, {Oman}  \& {White}}{{Sawala} et~al.}{2017}]{Sawala2017}
{Sawala} T.,  {Pihajoki} P.,  {Johansson} P.~H.,  {Frenk} C.~S.,  {Navarro}
  J.~F.,  {Oman} K.~A.,   {White} S.~D.~M.,  2017, \mn@doi [\mnras]
  {10.1093/mnras/stx360}, \href
  {http://adsabs.harvard.edu/abs/2017MNRAS.467.4383S} {467, 4383}

\bibitem[\protect\citeauthoryear{{Schaye} et~al.,}{{Schaye}
  et~al.}{2015}]{Eagle}
{Schaye} J.,  et~al., 2015, \mn@doi [\mnras] {10.1093/mnras/stu2058}, \href
  {http://adsabs.harvard.edu/abs/2015MNRAS.446..521S} {446, 521}

\bibitem[\protect\citeauthoryear{{Sellwood}}{{Sellwood}}{2013}]{Sellwood2013}
{Sellwood} J.~A.,  2013, {Dynamics of Disks and Warps}.
p.~923, \mn@doi{10.1007/978-94-007-5612-0_18}

\bibitem[\protect\citeauthoryear{{Springel}}{{Springel}}{2005}]{springel_2005}
{Springel} V.,  2005, \mn@doi [\mnras] {10.1111/j.1365-2966.2005.09655.x},
  \href {http://adsabs.harvard.edu/abs/2005MNRAS.364.1105S} {364, 1105}

\bibitem[\protect\citeauthoryear{{Springel} \& {Hernquist}}{{Springel} \&
  {Hernquist}}{2003}]{springel2003feedback}
{Springel} V.,  {Hernquist} L.,  2003, \mn@doi [\mnras]
  {10.1046/j.1365-8711.2003.06206.x}, \href
  {http://adsabs.harvard.edu/abs/2003MNRAS.339..289S} {339, 289}

\bibitem[\protect\citeauthoryear{{Springel} et~al.,}{{Springel}
  et~al.}{2008}]{springel2008}
{Springel} V.,  et~al., 2008, \mn@doi [\mnras]
  {10.1111/j.1365-2966.2008.14066.x}, \href
  {http://adsabs.harvard.edu/abs/2008MNRAS.391.1685S} {391, 1685}

\bibitem[\protect\citeauthoryear{{Stinson}, {Seth}, {Katz}, {Wadsley},
  {Governato}  \& {Quinn}}{{Stinson} et~al.}{2006}]{stinson2006}
{Stinson} G.,  {Seth} A.,  {Katz} N.,  {Wadsley} J.,  {Governato} F.,   {Quinn}
  T.,  2006, \mn@doi [\mnras] {10.1111/j.1365-2966.2006.11097.x}, \href
  {http://adsabs.harvard.edu/abs/2006MNRAS.373.1074S} {373, 1074}

\bibitem[\protect\citeauthoryear{Thornton \& Marion}{Thornton \&
  Marion}{2008}]{ThorntonAndMarion}
Thornton S.~T.,  Marion J.~B.,  2008, "Classical Dynamics of Particles and
  Systems", 5 edn.
"Cengage Learning"

\bibitem[\protect\citeauthoryear{{Torres-Flores}, {Epinat}, {Amram}, {Plana}
  \& {Mendes de Oliveira}}{{Torres-Flores} et~al.}{2011}]{TullyFisherModern}
{Torres-Flores} S.,  {Epinat} B.,  {Amram} P.,  {Plana} H.,   {Mendes de
  Oliveira} C.,  2011, \mn@doi [\mnras] {10.1111/j.1365-2966.2011.19169.x},
  \href {http://adsabs.harvard.edu/abs/2011MNRAS.416.1936T} {416, 1936}

\bibitem[\protect\citeauthoryear{{Vogelsberger}, {Genel}, {Sijacki}, {Torrey},
  {Springel}  \& {Hernquist}}{{Vogelsberger} et~al.}{2013}]{IllustrisFeedback}
{Vogelsberger} M.,  {Genel} S.,  {Sijacki} D.,  {Torrey} P.,  {Springel} V.,
  {Hernquist} L.,  2013, \mn@doi [\mnras] {10.1093/mnras/stt1789}, \href
  {http://adsabs.harvard.edu/abs/2013MNRAS.436.3031V} {436, 3031}

\bibitem[\protect\citeauthoryear{{Widrow}, {Pym}  \& {Dubinski}}{{Widrow}
  et~al.}{2008}]{WPDGalactICSReference}
{Widrow} L.~M.,  {Pym} B.,   {Dubinski} J.,  2008, \mn@doi [\apj]
  {10.1086/587636}, \href {http://adsabs.harvard.edu/abs/2008ApJ...679.1239W}
  {679, 1239}

\bibitem[\protect\citeauthoryear{{Yurin} \& {Springel}}{{Yurin} \&
  {Springel}}{2014}]{YurinSpringelGalic}
{Yurin} D.,  {Springel} V.,  2014, {GALIC: Galaxy initial conditions
  construction}, Astrophysics Source Code Library (\mn@eprint {ascl}
  {1408.008})

\bibitem[\protect\citeauthoryear{{Yurin} \& {Springel}}{{Yurin} \&
  {Springel}}{2015}]{YurinSpringelStellarDisks}
{Yurin} D.,  {Springel} V.,  2015, \mn@doi [\mnras] {10.1093/mnras/stv1454},
  \href {http://adsabs.harvard.edu/abs/2015MNRAS.452.2367Y} {452, 2367}

\bibitem[\protect\citeauthoryear{{Zemp}, {Gnedin}, {Gnedin}  \&
  {Kravtsov}}{{Zemp} et~al.}{2012}]{Zemp2012}
{Zemp} M.,  {Gnedin} O.~Y.,  {Gnedin} N.~Y.,   {Kravtsov} A.~V.,  2012, \mn@doi
  [\apj] {10.1088/0004-637X/748/1/54}, \href
  {http://adsabs.harvard.edu/abs/2012ApJ...748...54Z} {748, 54}

\bibitem[\protect\citeauthoryear{{de Boer}, {Belokurov}  \& {Koposov}}{{de
  Boer} et~al.}{2017}]{de_boer_et_al_2017}
{de Boer} T.~J.~L.,  {Belokurov} V.,   {Koposov} S.~E.,  2017, preprint, \href
  {http://adsabs.harvard.edu/abs/2017arXiv170609468D} {} (\mn@eprint {arXiv}
  {1706.09468})

\bibitem[\protect\citeauthoryear{{van Dokkum} et~al.,}{{van Dokkum}
  et~al.}{2013}]{van_dokkum_etal_2013}
{van Dokkum} P.~G.,  et~al., 2013, \mn@doi [\apjl]
  {10.1088/2041-8205/771/2/L35}, \href
  {http://adsabs.harvard.edu/abs/2013ApJ...771L..35V} {771, L35}

\makeatother
\end{thebibliography}




\newpage
\appendix

\section{Euler's Equations in Comoving Coordinates} \label{sec:derivation}

The time-evolution of the angular momentum vector ${\bf L}$ of a rigid
body acted upon by a torque $\boldsymbol{\tau}$ is given by

\begin{equation}
  \left(\frac{{d} \bf{L}}{{d} t}\right)_f = \left(\frac{{d} \bf{L}}{{d} t}\right)_b  + \boldsymbol \omega \times \bf{L}
  = \boldsymbol{\tau}
\end{equation}

\noindent where the subscripts $f$ and $b$ denote the frame of the
simulation box and the body frame, respectively.  In physical
coordinates, ${\bf L} = {\bf r \times p}$.  Alternatively, we can
write ${\bf L} = {\bf s\times q}$ where ${\bf s} = a^{-1}{\bf r}$
refer to comoving coordinates and ${\bf q} = a^2\dot{\bf s}$ is the
conjugate momentum to ${\bf s}$ (see \citet{QuinnIntegrators}).

For a rigid body, the components of the angular momentum are given by $L_i = I_{ij} \omega_j$ where $i,j$ run over $x,\,y,\,z$ and there is an implied sum over $j$.  Since \textsc{gadget-3} uses comoving coordinates, we write $I_{ij} = a^2 J_{ij}$ where $J$ is the moment of inertia tensor written in terms of the comoving coordinates, ${\bf s}$, rather than the physical coordinates, ${\bf r}$.  For convenience, we define a ``comoving'' angular velocity $\boldsymbol{\varpi} = a^{-2}\boldsymbol{\omega}$.  We then have $L_i = J_{ij} \varpi_j$.  Note that because of the symmetry of our disc, the moment of inertia tensor is diagonal with $J_{xx} = J_{yy}= J_{zz} \equiv J/2$.  The equations of motion for the Euler angles and the disc angular velocity are then given by the standard Euler equations of rigid body dynamics, modified to account for the time-dependence of the disc's moment of inertia:

\begin{equation}
\frac{d\phi}{dt} = a^{-2}\sin^{-1}{\theta} \left (\varpi_x\sin(\psi) + \varpi_y \cos(\psi)\right )~,
\end{equation}
\begin{equation}
\frac{d\theta}{dt} = a^{-2}\left (  \varpi_1\cos(\psi) - \varpi_y \sin(\psi)\right )~,
\end{equation}
\begin{equation}
J\frac{\varpi_x}{dt} + \varpi_x\frac{dJ}{dt}
+ J\varpi_y\varpi_z
=  \tau_x~,
\end{equation}
and
\begin{equation}
\frac{\varpi_y}{dt} + \varpi_y\frac{dJ}{dt}
- J\varpi_x\varpi_z
=  \tau_y~.
\end{equation}

\noindent We have omitted the equations for $\psi$ (rotations in the body frame about the symmetry axis) and $\varpi_z$ since these are determined directly from Eq.\,\ref{eq:omega3}.   


\bsp	
\label{lastpage}

\end{document}